\documentclass[reprint,twocolumn,superscriptaddress,secnumarabic,amssymb, nobibnotes, aps,pra]{revtex4-1}

\usepackage{amsmath}    % need for subequations
\usepackage{graphicx}    % need for figures
\usepackage{verbatim}    % useful for program listings
\usepackage{color}           % use if color is used in text
\usepackage{subfigure}   % use for side-by-side figures
\usepackage{hyperref}    % use for hypertext links, including those to external documents 
\usepackage{verbatim}  % and URLs
\definecolor{ao}{rgb}{0.0, 0.5, 0.0}
\usepackage[normalem]{ulem}

\begin{document}
\title{CaCu$_3$Ru$_4$O$_{12}$: a high Kondo-temperature transition metal oxide}
\author{D. Takegami}
\affiliation{Max Planck Institute for Chemical Physics of Solids, N{\"o}thnitzer Str. 40, 01187 Dresden, Germany}
\author{C. Y. Kuo}
\affiliation{Max Planck Institute for Chemical Physics of Solids, N{\"o}thnitzer Str. 40, 01187 Dresden, Germany}
\affiliation{National Synchrotron Radiation Research Center, 101 Hsin-Ann Road, 30076 Hsinchu, Taiwan}

\author{K. Kasebayashi}
\affiliation{Department of Physics and Electronics, Osaka Prefecture University 1-1 Gakuen-cho, Nakaku, Sakai, Osaka 599-8531, Japan}

\author{J.-G. Kim}
\affiliation{Department of Physics, POSTECH, Pohang 37673, Korea}

\author{C. F. Chang}
\affiliation{Max Planck Institute for Chemical Physics of Solids, N{\"o}thnitzer Str. 40, 01187 Dresden, Germany}
\author{C. E. Liu}
\affiliation{Max Planck Institute for Chemical Physics of Solids, N{\"o}thnitzer Str. 40, 01187 Dresden, Germany}
\affiliation{Department of Electrophysics, National Chiao Tung University, Hsinchu 300, Taiwan}

\author{C. N. Wu}
\affiliation{Max Planck Institute for Chemical Physics of Solids, N{\"o}thnitzer Str. 40, 01187 Dresden, Germany}
\affiliation{Department of Physics, National Tsing Hua University, Hsinchu 30013, Taiwan}

\author{D. Kasinathan}
\author{S. G. Altendorf}
\author{K. Hoefer}
\affiliation{Max Planck Institute for Chemical Physics of Solids, N{\"o}thnitzer Str. 40, 01187 Dresden, Germany}
\author{F. Meneghin}
\author{A. Marino}
\affiliation{Max Planck Institute for Chemical Physics of Solids, N{\"o}thnitzer Str. 40, 01187 Dresden, Germany}

\author{Y. F. Liao}
\author{K. D. Tsuei}
\author{C. T. Chen}
\affiliation{National Synchrotron Radiation Research Center, 101 Hsin-Ann Road, 30076 Hsinchu, Taiwan}
\author{K.-T. Ko}
\affiliation{Department of Physics, POSTECH, Pohang 37673, Korea}

\author{A. G{\"u}nther}
\affiliation{Experimental Physics V, University of Augsburg, 86135 Augsburg, Germany}
\author{S. G. Ebbinghaus}
\affiliation{Institute of Chemistry, Martin Luther University Halle-Wittenberg, 06120 Halle, Germany}

\author{J. W. Seo}
\author{D. H. Lee}
\author{G. Ryu}
\affiliation{Max Planck Institute for Chemical Physics of Solids, N{\"o}thnitzer Str. 40, 01187 Dresden, Germany}

\author{A. C. Komarek}
\affiliation{Max Planck Institute for Chemical Physics of Solids, N{\"o}thnitzer Str. 40, 01187 Dresden, Germany}

\author{S. Sugano}
\author{Y. Shimakawa}
\affiliation{Institute for Chemical Research, Kyoto University, Uji, Kyoto 611-0011, Japan}

\author{A. Tanaka}
\affiliation{Department of Quantum Matter, ADSM, Hiroshima University, Higashi-Hiroshima 739-8526, Japan}

\author{T. Mizokawa}
\affiliation{Department of Applied Physics, Waseda University, Shinjuku, Tokyo 169-8555, Japan}

\author{J. Kune{\v s}}
\affiliation{Institute of Solid State Physics, TU Wien, 1040 Vienna, Austria}
\author{L. H. Tjeng }
\altaffiliation{hao.tjeng@cpfs.mpg.de}
\affiliation{Max Planck Institute for Chemical Physics of Solids, N{\"o}thnitzer Str. 40, 01187 Dresden, Germany}
\author{A. Hariki}
\altaffiliation{hariki@pe.osakafu-u.ac.jp}
\affiliation{Department of Physics and Electronics, Osaka Prefecture University 1-1 Gakuen-cho, Nakaku, Sakai, Osaka 599-8531, Japan}
\affiliation{Institute of Solid State Physics, TU Wien, 1040 Vienna, Austria}

\date{\today}

\begin{abstract}
We present a comprehensive study of CaCu$_3$Ru$_4$O$_{12}$ using bulk sensitive hard 
and soft x-ray spectroscopy combined with local-density approximation (LDA) + dynamical 
mean-field theory (DMFT) calculations. Correlation effects on both the Cu and Ru ions can 
be observed. From the Cu $2p$ core level spectra we deduce the presence of magnetic 
Cu$^{2+}$ ions hybridized with a reservoir of itinerant electrons. The strong photon 
energy dependence of the valence band allows us to disentangle the Ru, Cu, and O 
contributions and thus to optimize the DMFT calculations. The calculated spin and charge 
susceptibilities show that the transition metal oxide CaCu$_3$Ru$_4$O$_{12}$ must be
classified as a Kondo system and that the Kondo temperature is in the range of 500-1000 K. 

\end{abstract}

\maketitle

Transition metal oxides show a wide variety of spectacular physical properties such as 
superconductivity, metal-insulator and spin-state transitions, unusually large magneto-resistance, 
orbital ordering phenomena, and multiferroicity \cite{cava00,khomskii14,keimer15}. Remarkably, 
heavy fermion or Kondo behavior is hardly encountered in oxides. While quite common in the 
rare-earth and actinide intermetallics \cite{Stewart84,Coleman06,pfleiderer09,wirth16}, one
may find perhaps only in the oxide LiV$_2$O$_4$ \cite{kondo97,urano00,shimoyamada06} 
indications for heavy fermion physics. 

\begin{figure}[h!]
    \includegraphics[width=0.90\columnwidth]{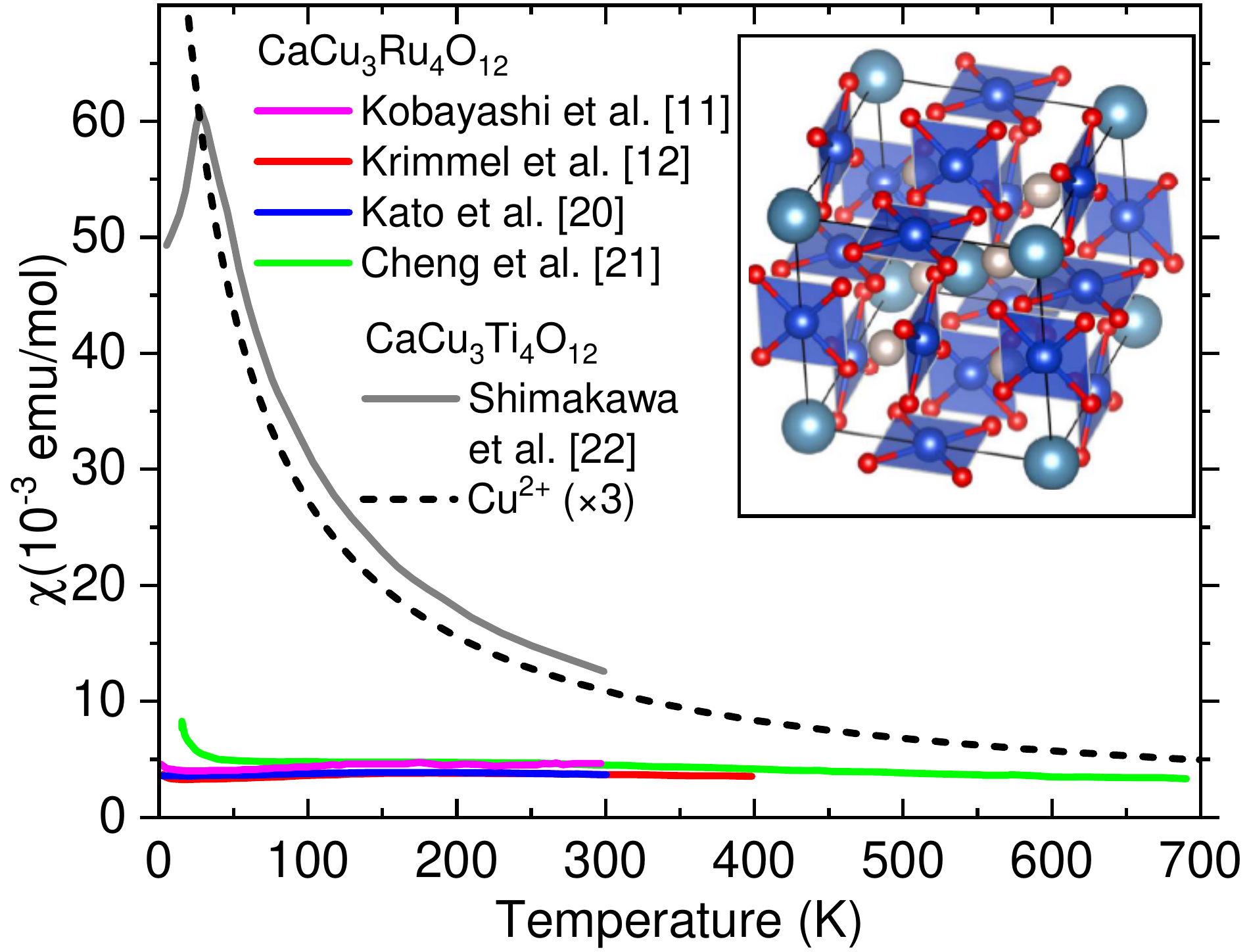}
    \caption{Magnetic susceptibility of CaCu$_3$Ru$_4$O$_{12}$ reproduced
    	from Refs. \cite{kobayashi04,Krimmel08,kato09,cheng13} and of CaCu$_3$Ti$_4$O$_{12}$ from
    	Ref. \cite{shimakawa08}. The black dashed line shows the Curie paramagnetic behavior for
    	a $S = 1/2$ Cu$^{2+}$ ion scaled by a factor of three. Inset: Crystal structure of 
    	CaCu$_3$Ru$_4$O$_{12}$ visualized by VESTA~\cite{vesta}.
		The blue, red, gray, and indigo blue spheres represent Cu, O, Ru, and Ca atoms, respectively.}
    \label{fig_chi}
\end{figure}

The discovery of the transition metal oxide CaCu$_3$Ru$_4$O$_{12}$ (CCRO) showing Kondo-like
properties therefore created quite an excitement~\cite{kobayashi04,Krimmel08,Krimmel09,Kao17}\color{black}. 
The crystal structure of this A-site ordered perovskite is shown in the inset of 
Fig.~\ref{fig_chi}. However, the Kondo interpretation has also met fierce reservations. 
It has been argued that the specific heat coefficient $\gamma$ does not deviate much from 
the band structure value, suggesting a minor role of the electronic correlations~\cite{xiang07}. 
Other interpretations of the mass enhancement have been put forward~\cite{tanaka09-1,tanaka09-2}. 
Electron spectroscopy studies have also not converged on the position or even the presence of 
the putative Kondo peak \cite{sudayama09,hollmann13,Liu20}.

Here we address the CCRO problem from a different perspective. Figure~\ref{fig_chi} shows the 
magnetic susceptibility of CCRO together with that of CaCu$_3$Ti$_4$O$_{12}$ (CCTO), as 
reproduced from Refs. \cite{kobayashi04,Krimmel08,kato09,cheng13} and Ref. \cite{shimakawa08}, respectively.
One can observe that CCTO follows, far above its 25 K N\'eel temperature, almost a text-book Curie-Weiss 
law that can be understood in terms of paramagnetic $S = 1/2$ Cu$^{2+}$ ions. By contrast, one 
can also see that CCRO shows a completely different behavior with a magnetic susceptibility that is an order of magnitude smaller. There are also no indications at all for magnetic order. 
If one believes that the system is non-magnetic and that Kondo physics does not take place 
in CCRO, then the Cu ions in CCRO have to be monovalent with the non-magnetic 
full-shell $3d^{10}$ configuration
or trivalent, which can be a nonmagnetic band insulator like NaCuO$_2$.

We show here 
that the Cu ions are definitely 
divalent and thus have spin degree of freedom. We have carried out a detailed photoelectron 
spectroscopy study using a wide range of photon energies in order to establish the presence 
of correlation effects on both the Cu and Ru ions as well as to disentangle the Ru, Cu, and O 
contributions to the valence band. This allowed us to tune the double counting corrections
in the LDA+DMFT calculations accordingly, making these calculations predictive for the low energy 
physics. We then were able to determine how the Cu$^{2+}$ magnetic moments can be screened. 
In particular, we will show that, in going from high to low temperatures, this screening takes 
place already at 500-1000 K, and that we thus must classify CCRO as a Kondo system with a very 
high Kondo temperature.  

Hard x-ray photoemission (HAXPES) measurements were carried out at the Max-Planck-NSRRC HAXPES 
end-station \cite{weinen15} at the Taiwan undulator beamline BL12XU of SPring-8 in Japan. The 
photon energy was set to $h\nu=6.5$~keV and the overall energy resolution was $\approx$ 270~meV 
as determined from the Fermi cutoff of a gold reference sample. Soft x-ray (resonant) photoelectron 
(PES) and absorption (XAS) spectroscopy experiments were performed at the NSRRC-MPI TPS 45A Submicron 
Soft x-ray Spectroscopy beamline at the Taiwan Photon Source (TPS) in Taiwan. The overall energy resolution 
when using 1.2~keV, 931~eV, and 440~eV photons was $\approx$ 150~meV, 125~meV, and 60~meV, respectively. 
Photoemission measurements in the vicinity of the Ru $4d$ Cooper minimum, i.e. at photon energies 
of $200$~eV, $150$~eV, and $100$~eV, were performed at the PLS-II 4A1 micro-ARPES beamline of the 
Pohang Light Source (PLS) in Korea. The overall energy resolution was $\approx$ 55~meV. Polycrystalline 
samples of CCRO were synthesized by solid-state reactions~\cite{ebbinghaus02}. Clean sample surfaces 
were obtained by cleaving sintered samples \textit{in situ} in ultra-high vacuum preparation 
chambers with pressures in the low $10^{-10}$~mbar range. The measurements at SPring-8 and TPS were 
carried out at $80$~K, and the measurements at PLS at $100$~K. We have used three different batches of 
CCRO samples for our spectroscopic measurements, all giving the same results, providing confidence
in the reliability of the data, see Appendix A.

Our calculations employ the LDA+DMFT scheme~\cite{metzner89,georges96,kotliar06}. We start with 
density functional calculations for the experimental crystal structure~\cite{Krimmel08}, see 
Fig.~\ref{fig_chi}, using the Wien2k code~\cite{wien2k} and construct the multi-band Hubbard 
model on the basis spanned by the Cu 3$d$, Ru 4$d$, and O 2$p$ Wannier functions obtained with 
the wannier90 package~\cite{wien2wannier,wannier90}. The on-site Coulomb interactions on the 
Cu and Ru sites are approximated with the density-density form with parameters 
($U$, $J$)=(8.5~eV, 0.98~eV) for Cu 3$d$ electrons and (3.1~eV, 0.7~eV) for Ru 4$d$ electrons, 
which are typical values for Cu and Ru systems\textcolor{black}{~\cite{petukhov03,hollmann13,gorelov10,pchelkina}}. 
The continuous-time quantum Monte Carlo (CT-QMC) method with the hybridization 
expansion~\cite{werner06,boehnke11,hafermann12} was used to solve the auxiliary Anderson impurity 
model (AIM). The double-counting corrections arising in LDA+X methods
\textcolor{black}{~\cite{karolak10,kotliar06}}, which fix the charge transfer energies on the Cu 
and Ru sites, were treated as adjustable parameters and their values fixed by comparison to the 
present valence band
\textcolor{black}{and core-level photoemission data as well as previous angle-resolved PES (ARPES) data.}
The valence spectra were obtained by analytic continuation of self-energy 
using the maximum entropy method~\cite{wang09,jarrell96}. The Cu 2$p$ and Ru $3d$ core-level XPS were 
calculated using the method of Refs.~\onlinecite{hariki17,hariki20,Ghiasi19}.

\begin{figure}
     \includegraphics[width=0.9\columnwidth]{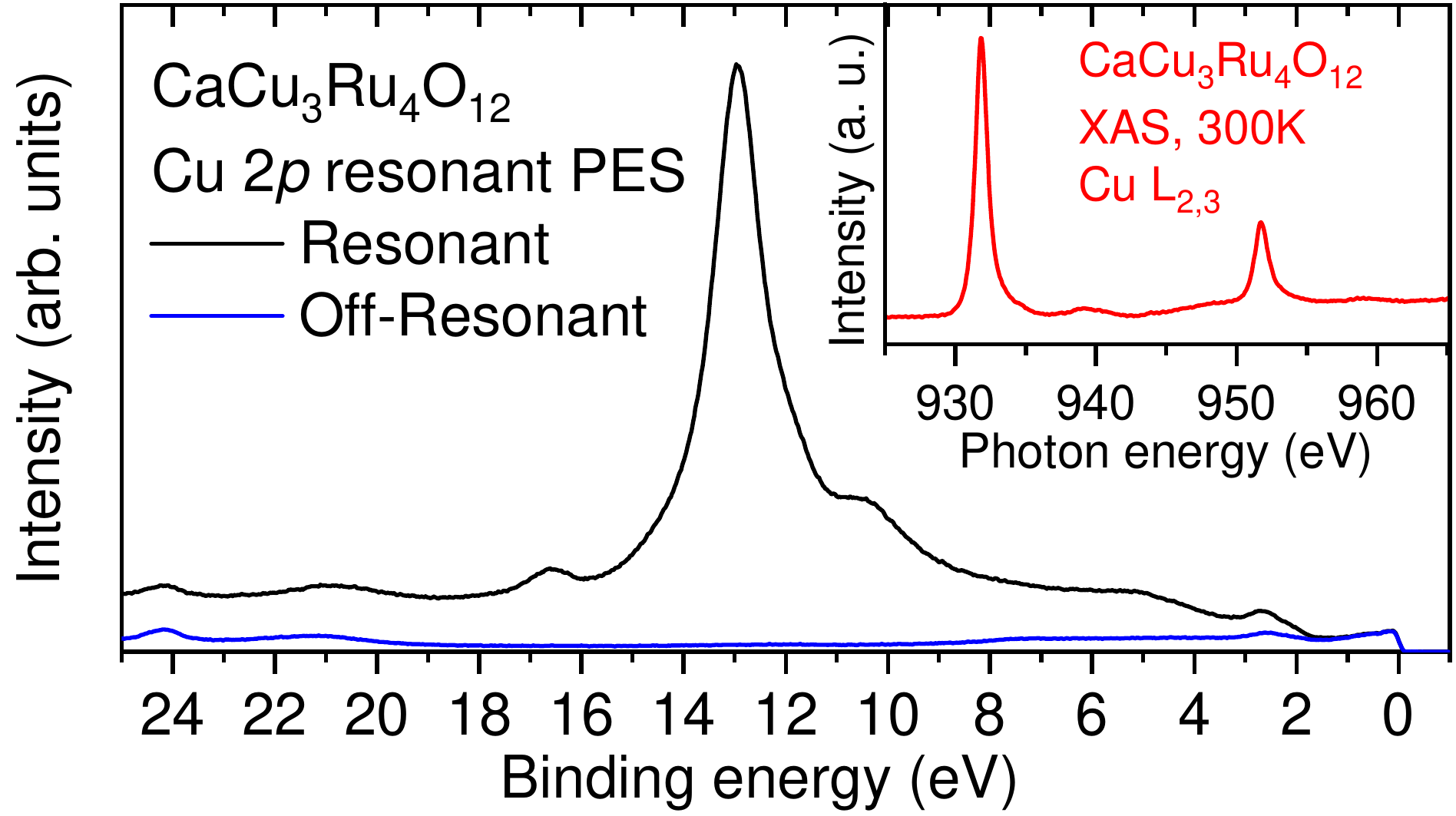}
    \caption{Valence band resonant photoemission of CaCu$_3$Ru$_4$O$_{12}$, 
     with the experimental spectra taken at the Cu $2p$ ($L_3$) resonance (h$\nu =931.2$ eV) 
     and at 10 eV below the resonance (h$\nu =921.2$ eV). The inset displays the experimental 
     Cu-$L_{2,3}$ x-ray absorption spectrum.}
    \label{fig:rpes}
\end{figure}

We have carried out XAS and valence band resonant PES measurements in the vicinity of the Cu $L_{2,3}$
edge, see Fig.~\ref{fig:rpes}. The peak positions and line shape of the spectra are characteristic for 
divalent Cu \cite{Tjeng1991,tjeng92,hollmann13, eskes90}. We can exclude that the Cu in CCRO is monovalent or trivalent since 
the spectral features of %Cu$^{1+}$
Cu$^{1+}$/Cu$^{3+}$
oxides are positioned at quite different energies \cite{tjeng92,hollmann13,Chin21}.
We thus can conclude that the Cu ions in CCRO possess a spin degree of freedom and that some form 
of screening must take place as to make their magnetic susceptibility to deviate dramatically from
the Curie-Weiss law.

\begin{figure}
	\includegraphics[width=0.8\columnwidth]{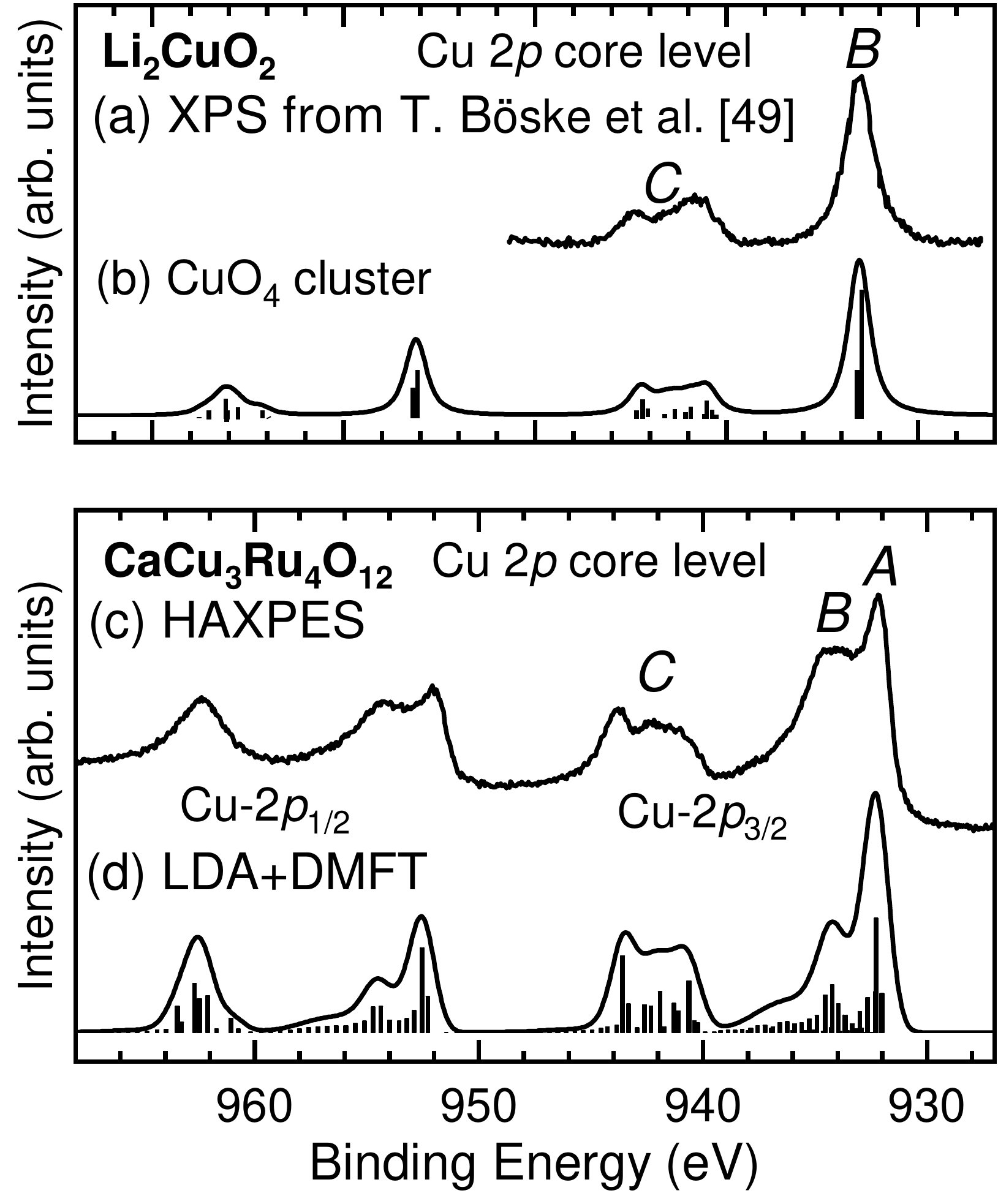}
	\caption{(a) Experimental Cu $2p$ core-level x-ray photoemission spectrum of Li$_2$CuO$_2$ 
		reproduced from Ref. \cite{Boeske98}.
		(b) Theoretical spectrum from the CuO$_4$ cluster model.
		(c) Experimental Cu $2p$ core-level HAXPES spectrum of CaCu$_3$Ru$_4$O$_{12}$.
		(d) Theoretical spectrum from the LDA+DMFT method.
	}
	\label{fig:core}
\end{figure}

Figure~\ref{fig:core} shows the Cu $2p$ core level HAXPES spectrum of CCRO together with that of 
Li$_2$CuO$_2$ as reproduced from Ref. \cite{Boeske98}. Here we took Li$_2$CuO$_2$ as a reference system 
which contains CuO$_4$ plaquettes that are weakly coupled 
\cite{Boeske98} and thus can 
serve for a comparison with CCRO which also contains rather isolated CuO$_4$ plaquettes, \textcolor{black}{see Fig.~\ref{fig_chi}}.
The spectra share the 
gross features: the main peak (B) at around 932-933 eV binding energy and the satellite 
(C) at 942 eV for the Cu $2p_{3/2}$ component. However, the fine structure differs considerably: the 
main peak of CCRO consists of two peaks (A and B) \cite{tran06,sudayama09} unlike the single peak (B) 
of Li$_2$CuO$_2$. 

The Cu $2p$ core level spectrum of Li$_2$CuO$_2$ is typical for a Cu$^{2+}$ oxide \cite{Ghijsen88}. 
Thanks to the weak coupling between the CuO$_4$ plaquettes, it can be explained accurately with 
a full multiplet single CuO$_4$-cluster calculation~\cite{tanaka94, param_cluster}, as shown in Fig.~\ref{fig:core}b. 
In contrast, the two-peak structure (A and B) of 
the CCRO main peak cannot be captured by the cluster model.
This is indicative of a screening process 
\cite{taguchi05,Panaccione08,tran06,sudayama09} which is present in CCRO but absent in Li$_2$CuO$_2$. 
\textcolor{black}{Since the CuO$_4$ plaquettes in CCRO are quite isolated from each other, a non-local screening mechanism 
%based on strong 
due to inter-Cu-cluster hopping is not expected to play an important role. We rather relate the screening process to the metallic state of CCRO.}
The LDA+DMFT calculations shown in Fig.~\ref{fig:core}d, reproduce the fine structure of 
the main peak very well. We thus indeed can infer that CCRO contains correlated magnetic Cu$^{2+}$ ions, which 
experience screening by conduction electrons. How strong or complete the screening is will be 
discussed below.  

%%%%%%%%%%%%%%%%%%%%%%%%%%%%%%%%%%%%%%%%%%%%
%%%%%%%%%%%%%%%%%%%%%%%%%%%%%%%%%%%%%%%%%%%%
\begin{figure}
	\includegraphics[width=0.98\columnwidth]{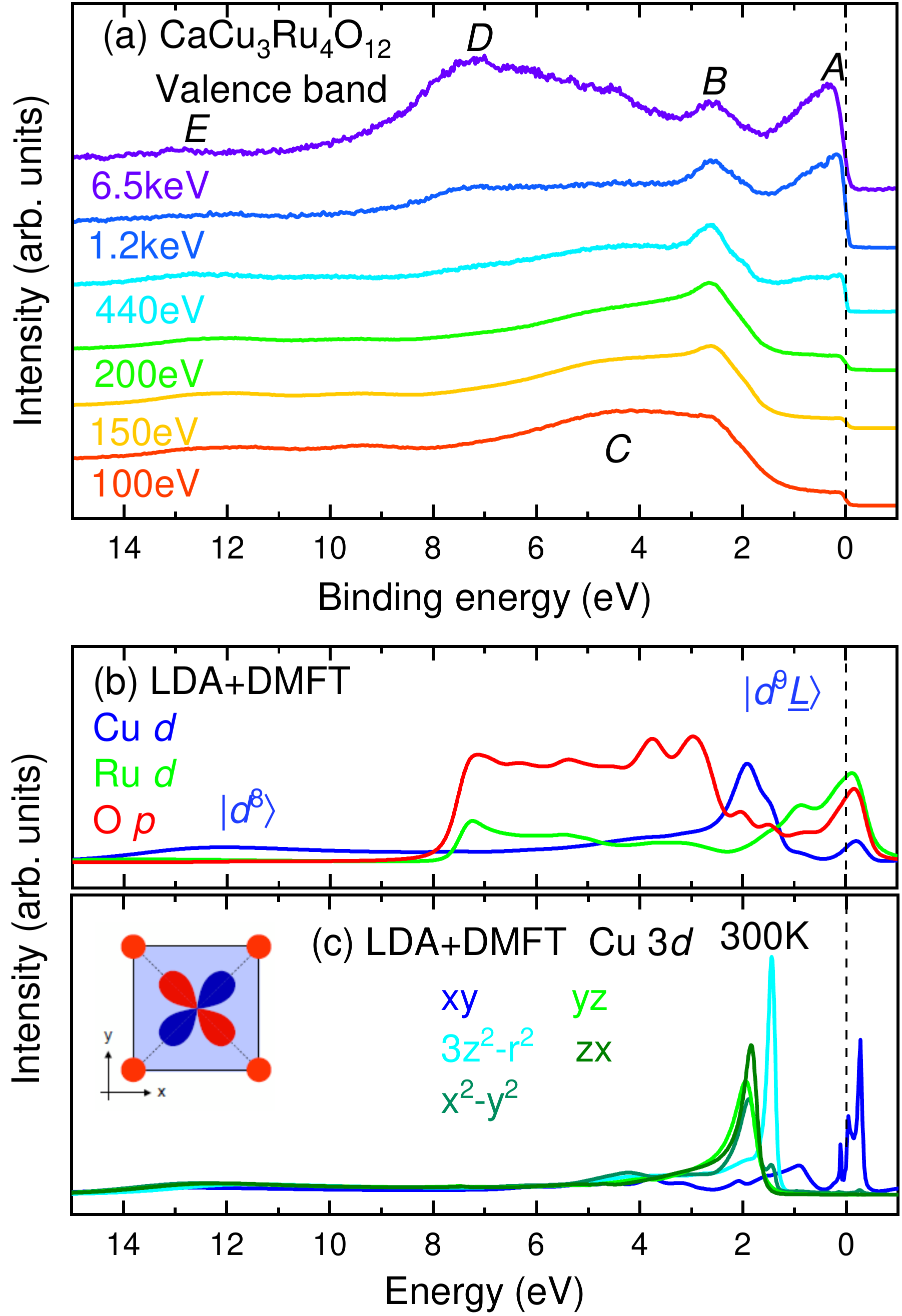}
	\caption{Valence band spectra of CaCu$_3$Ru$_4$O$_{12}$: 
		(a) experimental results measured at different photon energies,
		(b) LDA+DMFT spectral intensities for the Cu 3$d$, Ru 4$d$, and O 2$p$ states. The spectral broadening is taken into account using a 200~meV Gaussian to simulate the experimental resolution.
		(c) LDA+DMFT spectral intensities of the Cu $3d$ orbitals in CaCu$_3$Ru$_4$O$_{12}$.
		The Cu orbitals are defined in the local axis of the CuO$_4$ plane, as shown in the inset.}
	\label{fig:vb_nw}
\end{figure}
%%%%%%%%%%%%%%%%%%%%%%%%%%%%%%%%%%%%%%%%%%%%
%%%%%%%%%%%%%%%%%%%%%%%%%%%%%%%%%%%%%%%%%%%%

%%%%%%%%%%%%%%%%%%%%%%%%%%%%%%
\begin{figure}
	\includegraphics[width=0.7\columnwidth]{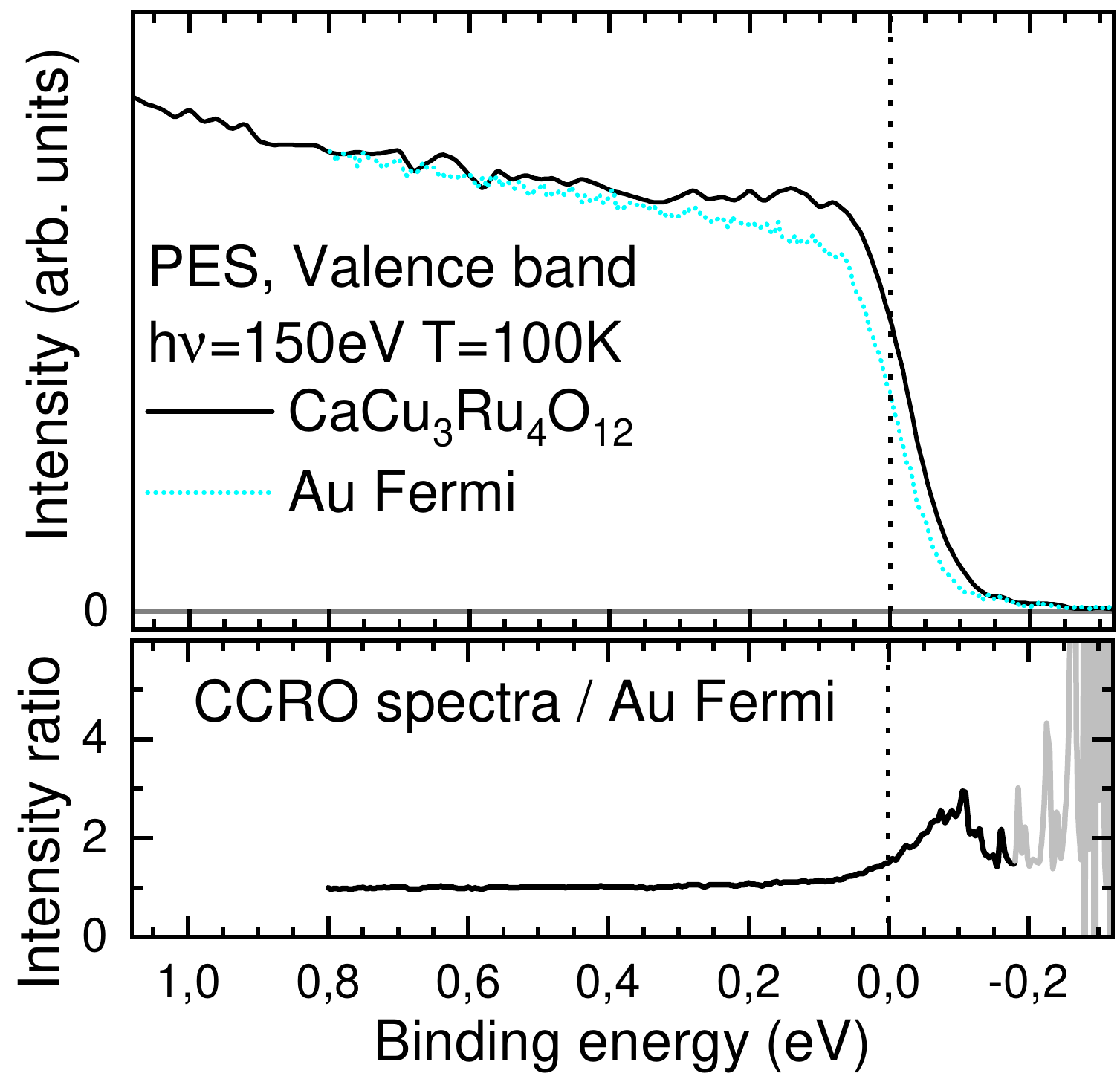}
	\caption{(top panel) Close-up of the Fermi level region of the CaCu$_3$Ru$_4$O$_{12}$ spectrum together with the gold spectrum.
		(bottom panel) Division of the CaCu$_3$Ru$_4$O$_{12}$ spectra by the gold spectrum.}
	\label{fig:div}
\end{figure}
%%%%%%%%%%%%%%%%%%%%%%%%%%%%%%%

%%%%%%%%%%%%%%%%%%
\begin{figure}
	\includegraphics[width=0.90\columnwidth]{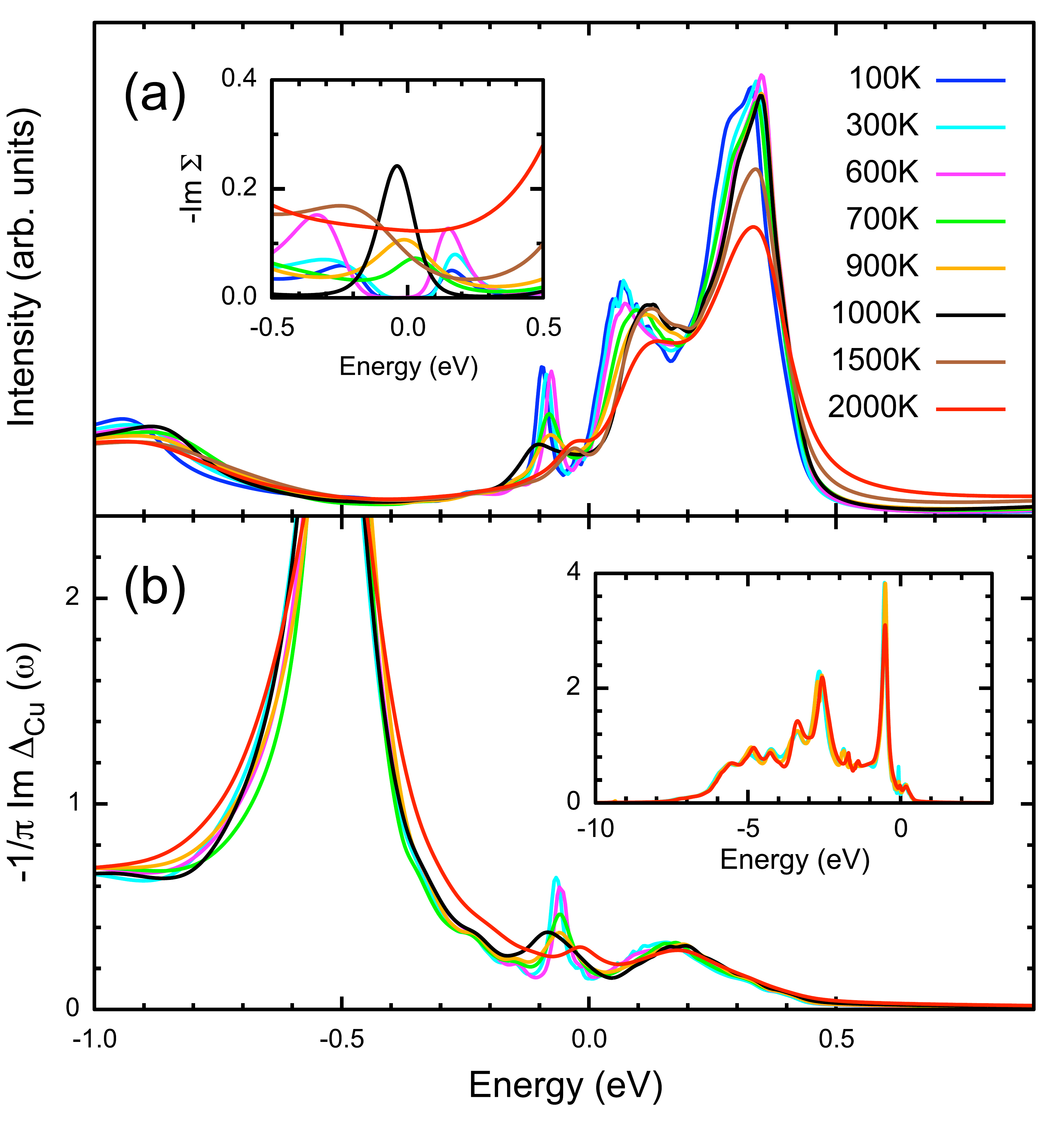}
	\caption{\textcolor{black}{(a) Temperature dependence of the Cu $xy$ spectral intensities in the LDA+DMFT method. The inset shows the imaginary part of the self-energy of the Cu $xy$ orbital. (b) LDA+DMFT hybridization densities $-1/\pi \ {\rm Im}\Delta(\omega)$ of the Cu $xy$ orbital. The inset shows the hybridisation densities in a wide energy range for selected temperatures (300~K, 900~K, 2000~K).}}
	\label{fig:cuxy}
\end{figure}
%%%%%%%%%%%%%%%%%

Figure~\ref{fig:vb_nw}a %(top) 
shows the experimental valence band spectra of CCRO \textcolor{black}{in a broad energy range} measured at various 
photon energies. Our motivation here is to make use of the different photon-energy dependence of the 
photoionization cross sections \cite{trzhaskovskaya01,trzhaskovskaya06, Yeh1985} to distinguish the Ru $4d$, 
Cu $3d$, and O $2p$ contributions to the spectra. The photon energy of 6.5~keV in HAXPES is much higher 
than the previously used photon energies of 1486.6~eV and~920 eV~\cite{tran06,sudayama09,hollmann13} 
and makes Ru $4d$ have the 
largest cross section relative to Cu $3d$ and O $2p$. The low photon energies of 200, 150, and 100 eV are close to the Cooper minimum of the Ru $4d$ cross section 
\cite{Yeh1985}, so that with these photon energies the Ru $4d$ signal gets maximally 
suppressed, enabling us to observe better the Cu $3d$ contribution. The ratio between the O $2p$ 
and Cu $3d$ cross sections also becomes continuously larger with lowering the photon energy 
\cite{Yeh1985}. See also Appendix B displaying the photon energy dependence of the cross 
sections in more detail.

In the set of valence band spectra, Fig.~\ref{fig:vb_nw}a, %(top panel), 
we can identify features 
labeled A, B, C, D, and E. Features A and D are very strong at 6.5 keV and diminish almost completely between 100-200 eV. 
This strongly suggests that 
A and D are derived from the Ru $4d$ orbitals. Features B and E are visible throughout the set, while the intensity of C is enhanced 
at 100 eV, the lowest photon energy of the set. This observation indicates that 
the features B and E are related to the Cu $3d$ while C is likely of O $2p$ origin. 
We point out that the positions of these five features do not change with the photon energy and coincide with the previously reported soft x-ray studies~\cite{tran06,sudayama09,hollmann13}.

The LDA+DMFT results are presented in Fig.~\ref{fig:vb_nw}b.
They corroborate the above
assignment.
The calculated Ru $4d$ spectrum matches the experimental features A and D, while the theoretical Cu 
$3d$ spectrum explains well the features B and E, the former being the Cu $|d^9\underline{L}\rangle$ 
and the latter the Cu $|d^8\rangle$ final state~\cite{Tjeng1991,tjeng92, hollmann13,eskes90}. The feature C 
is captured by the theoretical O $2p$ spectra. The calculations reveal that the spectrum around the 
Fermi level is dominated by hybridized Ru $4d$ and O $2p$ bands.
In Fig.~\ref{fig:vb_nw}c, the Cu $3d$ spectrum is decomposed into $xy$, 3$z^2$-$r^2$, $x^2$-$y^2$, $yz$, and $zx$ components. As shown in the inset, the Cu $3d$ $xy$ orbital points to the surrounding four oxygen sites. Therefore, the Cu $3d$ $xy$ orbital hybridize with the Ru $4d$ orbitals (via the O $2p$ orbitals) most strongly among the five Cu $3d$ orbitals.
\textcolor{black}{The Cu $3d$ $xy$ spectral density near the Fermi level is quite low  compared to the Ru $4d$ and O $2p$ one, see Fig.~\ref{fig:vb_nw}b, and most of its 
weight is 
above the Fermi level.}

 In order to experimentally detect the 
Cu contribution around 
the Fermi level, we focus on the larger peak at positive energies
and the spectra taken at 150~eV which is almost the photon energy to minimize the Ru $4d$ signal.
The top panel of Fig.~\ref{fig:div} 
displays a close-up of the spectra along with the corresponding gold reference spectrum taken under 
the same conditions. In order to look for the possible presence of states above the Fermi level, 
we divide the CCRO spectrum by the corresponding gold spectrum. The results are shown in the 
bottom panel of Fig.~\ref{fig:div}. We can identify clearly the presence of a sharp peak at about 
0.07-0.08 eV above the Fermi level, very consistent with the results of the LDA+DMFT calculations.

\textcolor{black}{Zooming to the vicinity of the Fermi level~\footnote{In order to reach lower temperatures we have used a model without the empty Ru $4d$ $e_g$ states here. The equivalence of Cu $3d$ $xy$ spectra in the models with and without
Ru $4d$ $e_g$ states is established in Appendix D.},
the calculated Cu $3d$ $xy$ spectrum in Fig.~\ref{fig:cuxy}a
reveals sharp temperature-dependent peaks.} 
\textcolor{black}{To put the present results in the context of Anderson impurity and
periodic Anderson models we show the Cu $3d$ $xy$ hybridization densities in 
Fig.~\ref{fig:cuxy}b. The global view in the inset shows a strongly asymmetric situation  with the Fermi energy located in the tail of the hybridization density.
The hybridization density exhibits only a minor temperature dependence. 
A sharp hybridization peak below the Fermi level, responsible for the peak around $-0.08$~eV 
in the Cu spectra also observed with angle-resolved PES (ARPES) around the $H$ point in the Brillouin zone~\cite{Liu20}, is an exception. This explains the 
different temperature behavior of the peaks in Fig.~\ref{fig:cuxy}a.
Damping of the peaks above the Fermi level with increasing temperature is not accompanied by changes of the hybridization function and thus reflects 
the Anderson/Kondo impurity physics controlled by the Kondo temperature.
Damping of the $-0.08$~eV peak arises from temperature-induced changes of the hybridization function, i.e., involves the feedback from the Cu ions (the temperature dependence of Ru contribution is negligible in the studied range). It is therefore
an Anderson/Kondo lattice effect related to somewhat lower coherence temperature~\cite{Burdin2000,Pruschke2000}.
}
The inset of  Fig.~\ref{fig:cuxy}a shows the inverse quasi-particle lifetime. The region of long lifetime (dip in the inverse lifetime) between -$0.1$~eV and $+0.05$~eV around the Fermi level found at lower temperatures, marks limits of the Fermi liquid theory.
\textcolor{black}{Between 600--700~K the scattering rate around Fermi level increases forming a peak in $\operatorname{Im}\Sigma$. Eventually (1500--2000~K) the scattering rate grows quasi-uniformly in entire low-energy regime (-0.5--0.5~eV) with increasing temperature.}

%%%%%%%%%%%%%%%%%%%
%%%%%%%%%%%%%%%%%%%
\begin{figure}
	\includegraphics[width=0.99\columnwidth]{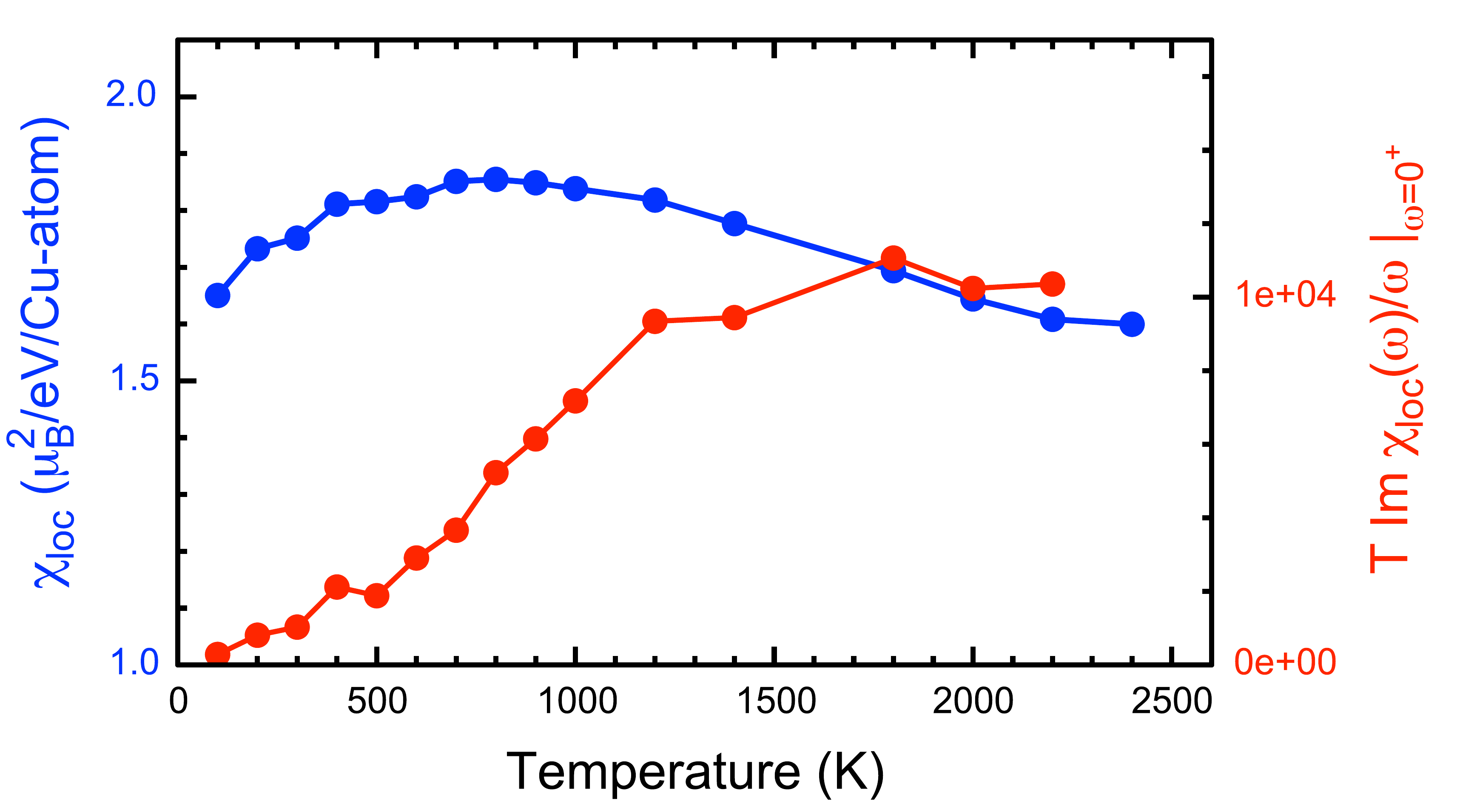}
	\caption{\textcolor{black}{Local susceptibility $\chi_{\rm loc}(T)$ (blue) 
	 and the dynamical spin susceptibility $\chi_{\rm loc}(\omega={+0})$ (red), calculated by the LDA+DMFT method.}
	 }
	\label{fig_loc}
\end{figure}
%%%%%%%%%%%%%%%%%%%
%%%%%%%%%%%%%%%%%%%
\begin{comment}
Having established that the LDA+DMFT calculations can reproduce very well the experimentally 
observed Ru $4d$, Cu $3d$, and O $2p$ contributions to the spectral weight, we now analyze the 
orbital character of the states closest to the Fermi level and their behavior with temperature.
Figure~\ref{fig:vb_nw}c (bottom panel) shows the orbitally resolved Cu $3d$ spectral weights 
over a wide energy range. A closer look reveals that the peak just above the Fermi level stems 
overwhelmingly from the $xy$ orbital, i.e. the Cu $3d$ orbital which is $\sigma$-bonded with 
the oxygen, see inset in Fig.~\ref{fig:vb_nw}c (bottom panel). This is consistent with the 
notion that the hole in a square-planar coordinated Cu$^{2+}$ ion resides in the $\sigma$-bonded 
orbital \cite{Eskes88, eskes90}. The calculations show that the 
affinity state, 
i.e. the state that one reaches when an electron is added to the system, is very close to the 
chemical potential in CCRO. 
The height of the corresponding peak decreases with 
increasing temperature, very reminiscent of Ce and Yb $4f$ systems 
\cite{Bickers87, Patthey90, Tjeng93}.
\end{comment}

Next, we discuss the local spin susceptibility 
$\chi_{\rm loc}$ at the Cu site. The top panel of Fig.~\ref{fig_loc} displays the $\chi_{\rm loc}$ 
obtained with LDA+DMFT. 
It exhibits the Curie behavior at 
high temperatures, and turns into a broad maximum at lower temperatures which is characteristic 
for Kondo screening. The deviation from the Curie behavior starts around 1000~K, suggesting a 
relatively high Kondo scale $T_{K}$. The calculated $\chi_{\rm loc}$ reproduces quite well the experimental susceptibility~\cite{kobayashi04} although an exact 
match can not be expected between the local and the uniform spin susceptibility, due to the contribution of the itinerant Ru 4$d$ - O $2p$ states.
The dynamical local susceptibility $\chi_{\rm loc}(\omega)$
is, nevertheless, directly related to the spin-lattice relaxation rate $1/T_1$ measured in  nuclear magnetic resonance (NMR) experiments~\cite{kato09,Krimmel08}
%%%%%%%%%%%%%
\begin{equation}
\frac{1}{T_1} \propto 
T  \lim_{\omega \to +0}{\rm Im} \frac{\chi_{\rm loc}(\omega)}{\omega}. \notag
\end{equation}
%%%%%%%%%%%%%%%%
The calculated temperature dependence of $1/T_1$ is shown in Fig.~\ref{fig_loc}.
\textcolor{black}{It shows an approximately linear increase at low-temperatures, which flattens into a constant behavior at around 1000~K. }
It is well-known that the former is characteristic of a Fermi liquid, while the latter of a fluctuating local moment. 
For CCRO, the NMR experiment was performed up to 700~K so far, where the absence of 
the constant
behavior was brought up as an evidence against the Kondo physics in this compound. Our result however suggests that CCRO manifests its local moment signature in NMR above the reported temperature.
The corresponding spin--spin correlation functions $\chi_{\rm spin}(\tau)$, 
see Appendix C, reflect the presence of an instantaneous Cu 3$d$ moment at all temperatures, 
which rapidly disappears on a short-time scale at temperatures below $T_K$. In contrast, the 
charge correlations are temperature independent. Rapid charge fluctuations present at all 
temperatures rather reflect the Cu-O bonding, not the hybridization with the states in the 
vicinity of the Fermi level.

The presence of magnetic Cu ions immersed in an itinerant band leads indeed to the emergence of Kondo physics, as demonstrated by our LDA+DMFT results. 
An important aspect for the long standing discussions about CCRO is our finding that the Kondo temperature is quite high, namely at around 700~K (between 500 and 1000~K). 
We would like to note that the parameters and double-counting corrections in our LDA+DMFT calculations have been tuned as to reproduce the available experimental (HAX)PES 
and the recent ARPES~\cite{Liu20} spectra, 
\textcolor{black}{details can be found in Appendix~D.}
In this respect it is worth 
mentioning that we have also included on-site Coulomb interactions at the Ru site.
This was necessary to explain the presence of a satellite structure in the Ru $3d$ core level PES spectrum, see Appendix D.
It turned out that the inclusion of 
\textcolor{black}{$U_{\rm Ru}$}
in the Ru $4d$ shell has also a substantial influence
on the low energy properties of CCRO. 
We have calculated the local spin susceptibility $\chi_{\rm loc}$ at the Cu site with $U_{\text{Ru}}=3.1$~eV and $U_{\text{Ru}}=0.0$~eV,
\textcolor{black}{see Appendix~D,
indicating that
correlations on the Ru site influence the screening process on the Cu site.}
This can be traced back to the influence of the Ru $U$ on the shape of the Ru $4d$ - O $2p$ band. 
\textcolor{black}{In Appendix~D,}
we show the Cu $xy$ spectral intensities calculated with $U_{\rm Ru}=0$~eV on the Ru 4$d$ shell, where the agreement to the available PES data is surrendered.

The high Kondo temperature we have found from our LDA+DMFT calculations implies that the 
contribution of the Kondo screening process to the low temperature specific heat is modest, 
thus explaining why band structure calculations can seemingly reproduce the experimentally 
observed $\gamma$ value of the specific heat rather well since the main contribution comes 
from the uncorrelated Ru-O derived bands. The high Kondo temperature also implies that 
one needs to go to very high temperatures to see the appearance of local Cu moments,
as demonstrated in Fig.~\ref{fig_loc}.
We can thus infer that upon going from high to low temperatures the Kondo screening process is practically completed already at 300 K, and that lowering the temperature further would not produce significant changes in the electronic and magnetic properties. 
%So 
It is understandable 
%now 
that there has been a controversy concerning 
Kondo physics in CaCu$_3$Ru$_4$O$_{12}$ since its signatures 
in low temperature measurements are weak.

Nevertheless, 
Kondo physics is present, as evidenced by the low-temperature disappearance of the Cu$^{2+}$ magnetic susceptibility for which we were able to provide a quantitatively explanation using our LDA+DMFT calculations that include fine tuning of the parameters from a detailed comparison to bulk sensitive photoemission data. Our findings indicate that the material class CaCu$_3$M$_4$O$_{12}$ indeed provides a unique opportunity to explore Kondo phenomena in transition metal compounds, where one may achieve lower Kondo temperatures by suitably varying the M constituent.

The authors thank A. Sotnikov and J. Fern\'andez Afonso, Mathias Winder for fruitful discussions. 
A.H. and J.K. are supported by the European Research Council (ERC) under the European Union's 
Horizon 2020 research and innovation programme (grant agreement No.~646807-EXMAG). A.H. was supported by JSPS KAKENHI Grant Number 21K13884. The computations were performed at the Vienna Scientific Cluster (VSC). The research in Dresden 
is supported by the Deutsche Forschungsgemeinschaft (DFG) through Grant No. 320571839 and 
SFB 1143 (project-id 247310070). The experiments in Taiwan and Korea were facilitated by the 
Max Planck-POSTECH-Hsinchu Center for Complex Phase Materials. {\color{black}S.S. and Y.S. are supported by JSPS KAKENHI Grants-in-Aid for Scientific Research (No. 19H05823 and No. 20H00397) and by a JSPS Core-to-Core Program (A) Advanced Research Networks.}

\appendix

\section{Sample consistency}

\begin{figure}[h]
    \vspace*{0.4cm}
     \includegraphics[width=0.99\columnwidth]{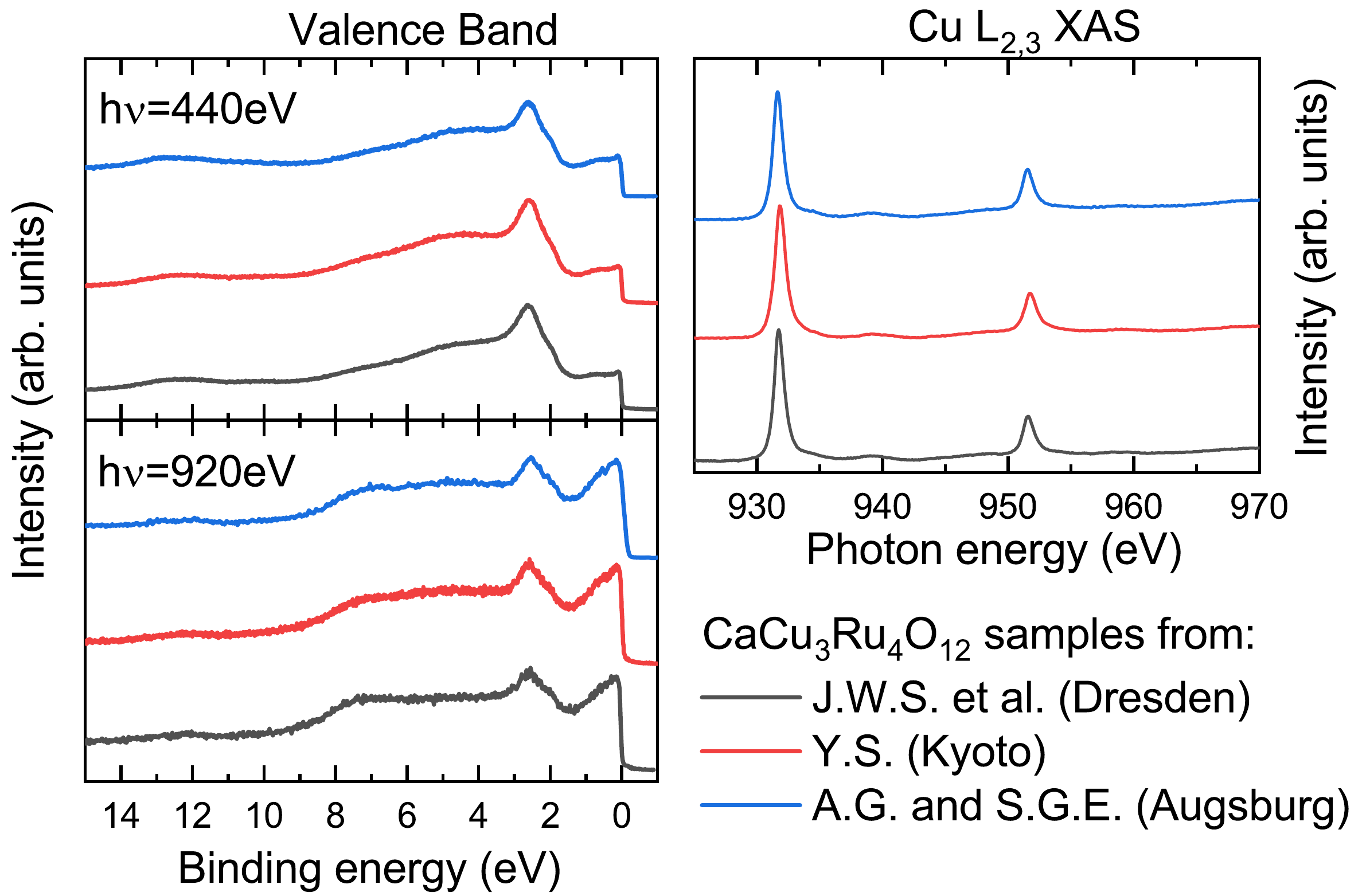}
    \caption{Valence band PES (left) and Cu-$L_{2,3}$ XAS (right) of CaCu$_3$Ru$_4$O$_{12}$ 
     samples synthesized by the different groups used in this work.}
    \label{fig:smp}
\end{figure}

In order to ensure that the experimental spectra and their features are intrinsic to the material, 
we have utilized three different batches of samples synthesized by three different groups: 
A.C. Komarek's group from Max Planck Institute for Chemical Physics of Solids, Dresden, 
A. G{\"u}nther from the University of Augsburg, and Y. Shimakawa's group from the Kyoto 
University. Figure \ref{fig:smp} shows the comparison of the valence band PES and XAS spectra 
taken from these three batches of samples. 

The results match perfectly, and thus confirm that the
data presented in this paper are not sample specific or due to extrinsic contributions

\section{Photoionization cross sections}

\begin{figure}
    \vspace*{0.4cm}
     \includegraphics[width=0.95\columnwidth]{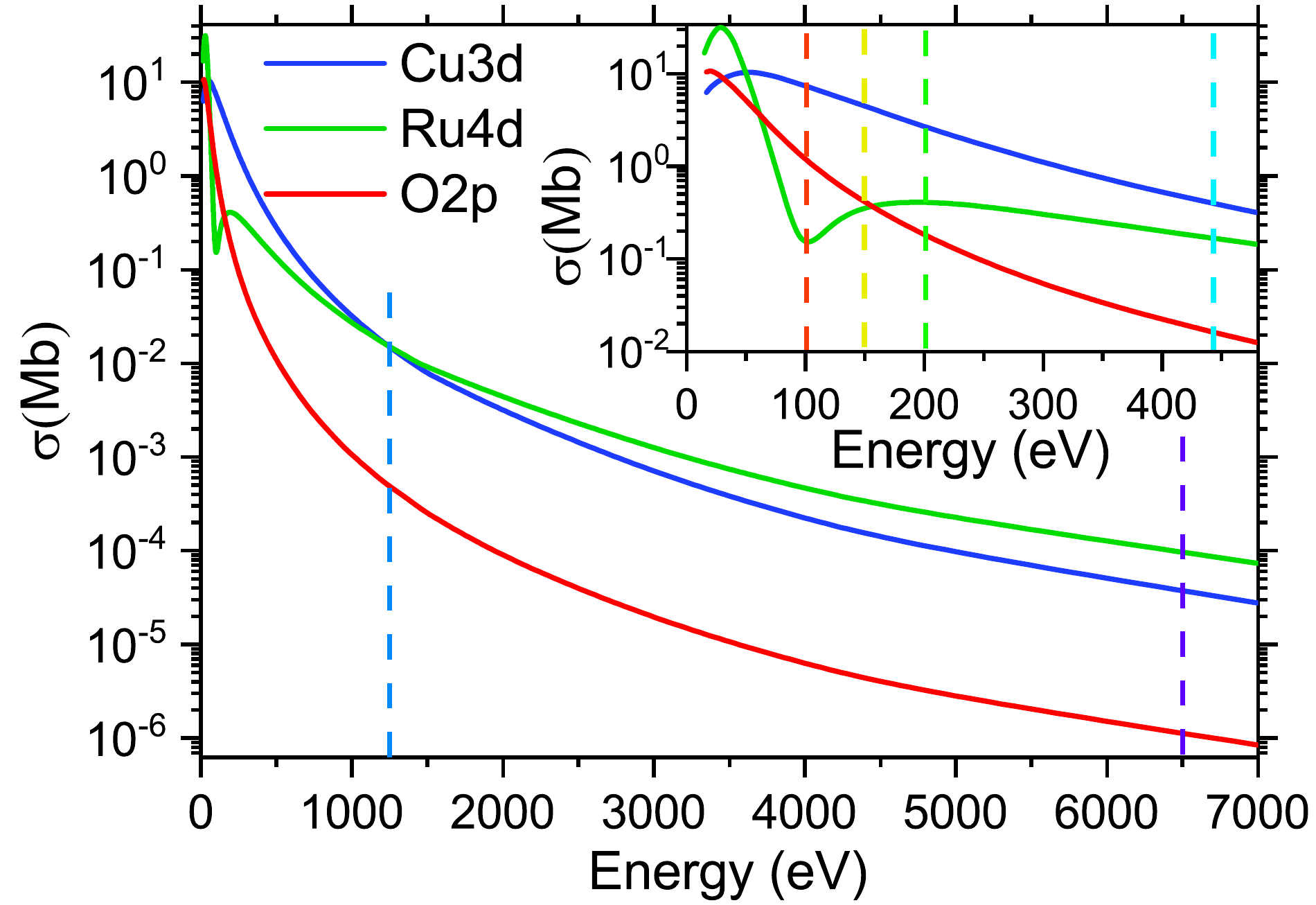}
    \caption{Photoionization cross section values of Cu $3d$, Ru $4d$, and O $2p$ 
      interpolated from the data tabulated in Refs. \cite{Yeh1985,trzhaskovskaya01,trzhaskovskaya06}. 
     The vertical lines indicate the photon energies used for the photoemission data shown in Fig. \ref{fig:vb_nw}.}
    \label{fig:cs}
\end{figure}

For the valence band of CaCu$_3$Ru$_4$O$_{12}$ the most relevant contributions are Cu $3d$, Ru $4d$, 
and O $2p$. Figure \ref{fig:cs} shows the photon energy dependence of the photoionization cross sections
as interpolated from the data tabulated in Refs. \cite{Yeh1985,trzhaskovskaya01,trzhaskovskaya06}. 
The vertical lines highlight the photon energies used in this study.  At high photon energies (HAXPES, 6.5 keV), 
the Ru $4d$ provides a larger signal than the Cu $3d$ and the O $2p$. Going to lower photon energies, the Cu $3d$
becomes gradually stronger, and with a crossover at $1.2$~keV, it becomes the dominant contributor. At around 
$100$~eV, there is a minimum in the Ru $4d$ cross section, also known as the Cooper minimum, 
which provides the ideal condition for studying the Cu $3d$ contributions. At this photon energy, the Cu $3d$ signal
is enhanced by a factor $\approx$ 70 with respect to the Ru $4d$. For lower photon energies, the Ru $4d$ quickly 
gains strength and becomes again the dominant contributor, making these low photon energies less useful when
searching for the Cu $3d$ signal close to the Fermi level. As for the O $2p$ contribution, we observe that it is highly 
suppressed for high energies, but becomes only competitive with the Cu $3d$ for energies below $\approx$ 50 eV.

\section{LDA+DMFT computational method}

\textcolor{black}{Below we describe the LDA+DMFT scheme~\cite{metzner89,georges96,kotliar06} employed to analyze the experimental data. We start with density functional calculations for the experimental crystal structure of CCRO~\cite{Krimmel08} using the Wien2k code~\cite{wien2k}. Then we construct the multi-band Hubbard model on the basis spanned by the Cu 3$d$, Ru 4$d$, and O 2$p$ Wannier functions from the LDA bands using wien2wannier and wannier90 packages~\cite{wien2wannier,wannier90}. The multi-band Hubbard model is augmented with the local electron-electron interaction within the Cu 3$d$ and Ru 4$d$ shells giving the Hamiltonian,
%%%%%%%%%%%%%%%%%%%%%%%%%%%%%%%%%%%%%%%%%%%%%%%%
\begin{eqnarray}
 H &=& \sum_{\textit{\textbf{k}}} 
 \begin{pmatrix}
      \textit{\textbf{c}}^{\dag}_\textit{\textbf{k}} & \textit{\textbf{r}}^{\dag}_\textit{\textbf{k}} & \textit{\textbf{p}}^{\dag}_\textit{\textbf{k}}
 \end{pmatrix}
 \begin{pmatrix}
     h^{cc}_{\textit{\textbf{k}}}-\mu_{\rm cr}^{\rm Ru} & h^{dp}_{\textit{\textbf{k}}} & h^{cp}_{\textit{\textbf{k}}} \\
     h^{rc}_{\textit{\textbf{k}}} & h^{rr}_{\textit{\textbf{k}}}-\mu_{\rm dc}^{\rm Cu} & h^{rp}_{\textit{\textbf{k}}} \\
     h^{pc}_{\textit{\textbf{k}}} & h^{pr}_{\textit{\textbf{k}}} & h^{pp}_{\textit{\textbf{k}}}
 \end{pmatrix}
   \begin{pmatrix}
      \textit{\textbf{c}}_\textit{\textbf{k}} \\
      \textit{\textbf{r}}_\textit{\textbf{k}} \\
      \textit{\textbf{p}}_\textit{\textbf{k}}
 \end{pmatrix} \notag \\
 &+& \sum_{i}W^{dd}_{i,{\rm Cu}}+\sum_{i}W^{dd}_{i,{\rm Ru}}. \notag
\end{eqnarray}
%%%%%%%%%%%%%%%%%%%%%%%%%%%%%%%%%%%%%%%%%%%%%%%%
Here, $\textit{\textbf{c}}^{\dag}_\textit{\textbf{k}}$ is an operator-valued vector whose elements are Fourier transforms of $c_{\gamma i}$, that annihilate the Cu 3$d$ electron in the orbital $\gamma$ in the $i$-th unit cell. Similarly $\textit{\textbf{r}}^{\dag}_\textit{\textbf{k}}$ and $\textit{\textbf{p}}^{\dag}_\textit{\textbf{k}}$ are those for Ru 4$d$ and O 2$p$ electrons. The on-site Coulomb interaction $W_{i,{\rm Cu}}^{dd}$ and $W_{i,{\rm Ru}}^{dd}$ on Cu and Ru sites is approximated with the density-density form with parameters ($U$, $J$)=(8.5~eV, 0.98~eV) for Cu 3$d$ electrons and (3.1~eV, 0.7~eV) for Ru 4$d$ electrons, which are typical values for Cu and Ru systems~\cite{petukhov03,hollmann13,gorelov10,pchelkina}. The double-counting terms $\mu_{\rm dc}^{\rm Ru}$, $\mu_{\rm dc}^{\rm Cu}$, which correct for the $d$--$d$ interaction present in the LDA step~\cite{karolak10,kotliar06}, renormalize the $p$-$d$ splitting and thus the charge-transfer energy.
\textcolor{black}{We have fixed the double-counting values to 
$\mu_{\rm dc}^{\rm Cu}=70.1$~eV and $\mu_{\rm dc}^{\rm Ru}=16.2$~eV ($\mu_{\rm dc}^{\rm Ru}=13.4$~eV for the model without 
Ru $e_g$ states) by comparison to the
photoemission spectroscopy data discussed in Appendix~D.}
The CT-QMC
method with the hybridization expansion~\cite{werner06,boehnke11,hafermann12,hariki15} was used to solve the auxiliary Anderson impurity model (AIM) in the DMFT self-consistent calculation.
\textcolor{black}{The valence spectra were obtained by analytic continuation of self-energy $\Sigma(\varepsilon)$ using the maximum entropy method~\cite{wang09,jarrell96}.
The hybridization function $\Delta(\varepsilon)$ for the Cu 3$d$ orbital $\gamma$ ($|d_{\gamma}\rangle$) is given by~\cite{georges96,kotliar06}}
%%%%%%%%%%%%%%%%%%%%%%%%%%%%%%%%%%%%%%%%%%%%%%%%
\begin{eqnarray}
    \Delta_{\gamma}(\varepsilon)=\langle d_{\gamma}|\varepsilon-h_0-\Sigma(\varepsilon)-G^{-1}(\varepsilon)|d_{\gamma}\rangle,  \notag
\end{eqnarray}
%%%%%%%%%%%%%%%%%%%%%%%%%%%%%%%%%%%%%%%%%%%%%%%%
\textcolor{black}{where $G(\varepsilon)$ and $h_0$ are the local Green's function and the one-body part of the on-site Hamiltonian at the Cu site, respectively. $\Delta (\varepsilon)$ does not depend on the spin, and small off-diagonal elements between different orbitals are neglected.}
The Cu 2$p$ core-level spectrum was calculated using the method of Refs.~\onlinecite{hariki17,hariki20}, where the AIM with the DMFT hybridization density is extended to include the Cu 2$p$ core orbitals~\cite{hariki17,Ghiasi19}. The configuration interaction scheme with 25 bath states representing the DMFT hybridization density is employed to evaluate the 2$p$ spectra.}

\begin{figure}
    \includegraphics[width=0.98\columnwidth]{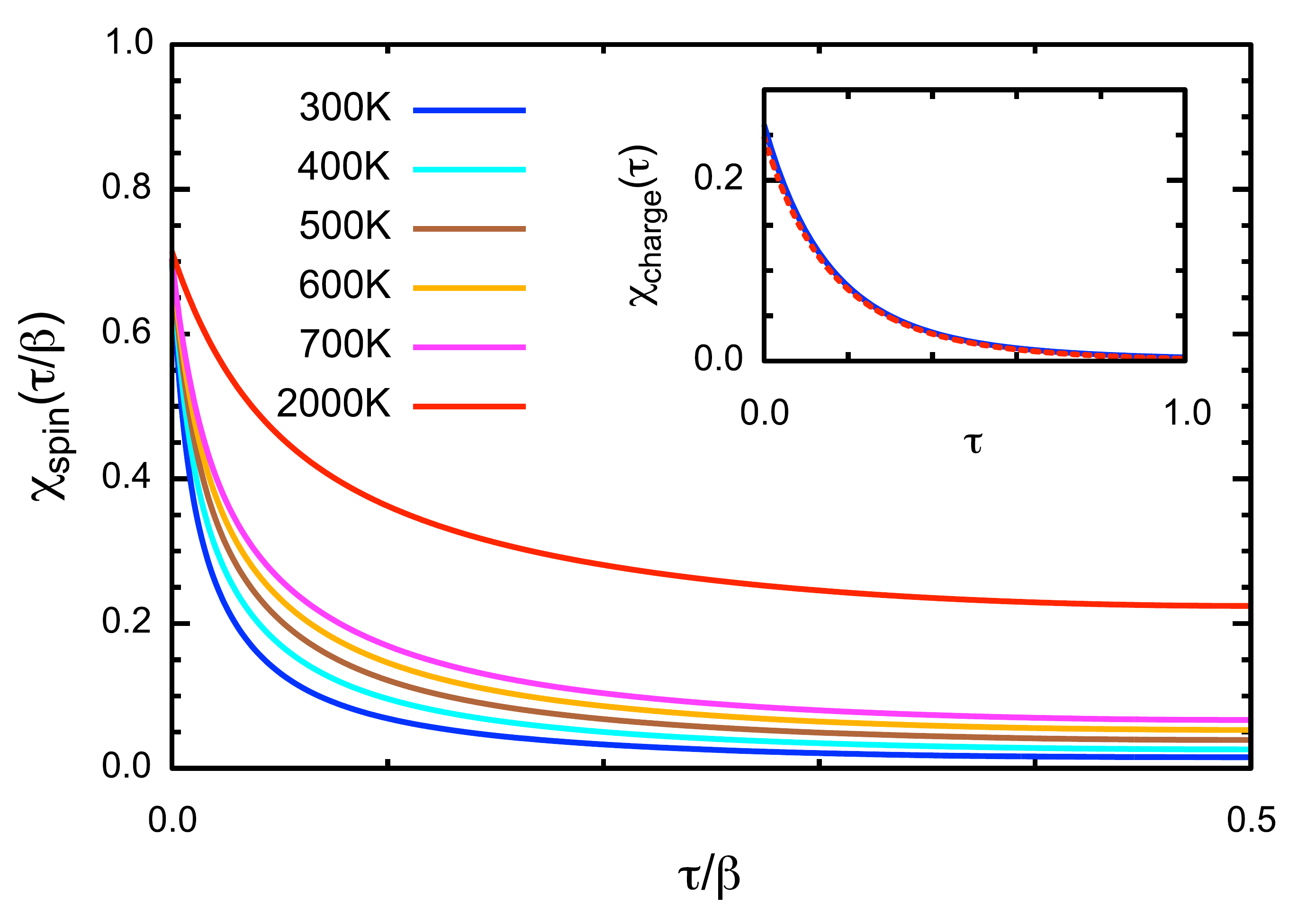}
    \caption{Correlation function $\chi (\tau)$ of the spin and charge (inset) channel calculated by the LDA+DMFT method, where $\beta$=1/${k_{\rm B}T}$ is the inverse temperature.}
    \label{fig_loctau}
\end{figure}

Figure~\ref{fig_loctau} shows the local spin $\chi_{\rm spin}(\tau)$ and charge $\chi_{\rm charge}(\tau)$ correlation functions at the Cu site. $\chi_{\rm spin}(\tau)$ is given as $\chi_{\rm spin}(\tau)=\langle \hat{m}_z(\tau)\hat{m}_z(0)\rangle$~\cite{krapek12,hariki17ru} and $\chi_{\rm charge}(\tau)$ is given by $\chi_{\rm charge}(\tau)=\langle \delta \hat{n}(\tau)\delta \hat{n}(0)\rangle$~\cite{Ylvisaker09}, where $\delta \hat{n}=\hat{n}-\langle \hat{n}\rangle$ with the Cu $d$-occupation number operator $\hat{n}$. These quantities are calculated by the impurity Anderson model with the DMFT hybridization using the CT-QMC method. The local susceptibility $\chi_{\rm loc}(T)$ is obtained by integrating the $\chi_{\rm spin}(\tau)$ at temperature $T$ with respect to the imaginary time $\tau$,
%%%%%%%%%%%%%%%%%%%%%%%%%%%%%%%%%%%%%%%%%%%%%%%%%
\begin{align}
\chi_{\rm loc}(T)&=\int_{0}^{1/T}d\tau \langle m_z(\tau)m_z(0)\rangle. \notag 
\end{align}
%%%%%%%%%%%%%%%%%%%%%%%%%%%%%%%%%%%%%%%%%%%%%%%%
A rigid instantaneous Cu 3$d$ spin moment is present for all temperatures, as imprinted in the temperature-independent value of $\chi_{\rm spin}(\tau=0)$, see Fig.~\ref{fig_loctau}. The spin moment survives on a long-time scale at high temperatures (see e.g.~at 2000~K), giving the Curie behavior in $\chi_{\rm loc}$ shown in the main text, while it disappears on short timescales at low temperatures due to the Kondo screening by the Ru 4$d$ - O $2p$ bands, giving the Pauli-like behavior in $\chi_{\rm loc}$.  In contrast to the temperature dependence in $\chi_{\rm spin}(\tau/\beta)$, the local charge correlation function $\chi_{\rm charge}$ is substantially suppressed at all temperatures, see the inset of Fig.~\ref{fig_loctau}, indicating the frozen charge fluctuation of the Cu 3$d$ electrons in CCRO, as expected in the Kondo regime. Thus the calculated spin and charge correlation functions support the Kondo behavior of Cu 3$d$ electrons in the studied material. We point out that the $\chi_{\rm spin}(\tau=0)$ value is reduced to 0.7 from 1.0 which corresponds to the $S=1/2$ Kondo limit. The reduction comes from the mixture of the $|d^{10}\underline{L}\rangle$ configuration by the Cu-O hybridization, where $|\underline{L}\rangle$ represents an O 2$p$ hole, but it does not show a temperature dependence. This behavior characterizes the effect of charge-transfer in the Cu-O subsystem. The energy scale of the Kondo 
physics therefore builds on the ligand-metal hybridization in addition to the Coulomb interaction $U$ at the Cu site.

%%%%%%%%%%%%%%%%%%%%%%%%%%%%%%%%%%%%%%%%%%%%%%%%%%%%%%%%%%%%%%%%%%%%%%%%%%%%%%%%%%%%%%%%%%%%%%%%%
%%%%%%%%%%%%%%%%%%%%%%%%%%%%%%%%%%%%%%%%%%%%%%%%%%%%%%%%%%%%%%%%%%%%%%%%%%%%%%%%%%%%%%%%%%%%%%%%%
\section{\textcolor{black}{Parameters in the LDA+DMFT simulation}}
\textcolor{black}{In the calculations we used two versions of the model described in Appendix C. First model contains
all Ru $4d$ states, while the second contains only the Ru $t_{2g}$ states present at and below the Fermi level. 
The first model, which we consider as the primary one, is needed to capture the full spectrum, in particular the feature 
D of Fig.~\ref{fig:vb_nw}a. The second model is used for numerical convenience as it allows us to perform LDA+DMFT calculations at lower temperature \textcolor{black}{(below 300~K)}. 
\textcolor{black}{It is employed to compute the low-energy Cu spectral intensities (Fig.~\ref{fig:cuxy}) and response functions (Fig.~\ref{fig_loc}) including temperatures below 300~K. Its results are validated by comparison to the first model at the higher temperatures.} 
}

\begin{figure}[t]
	\includegraphics[width=0.95\columnwidth]{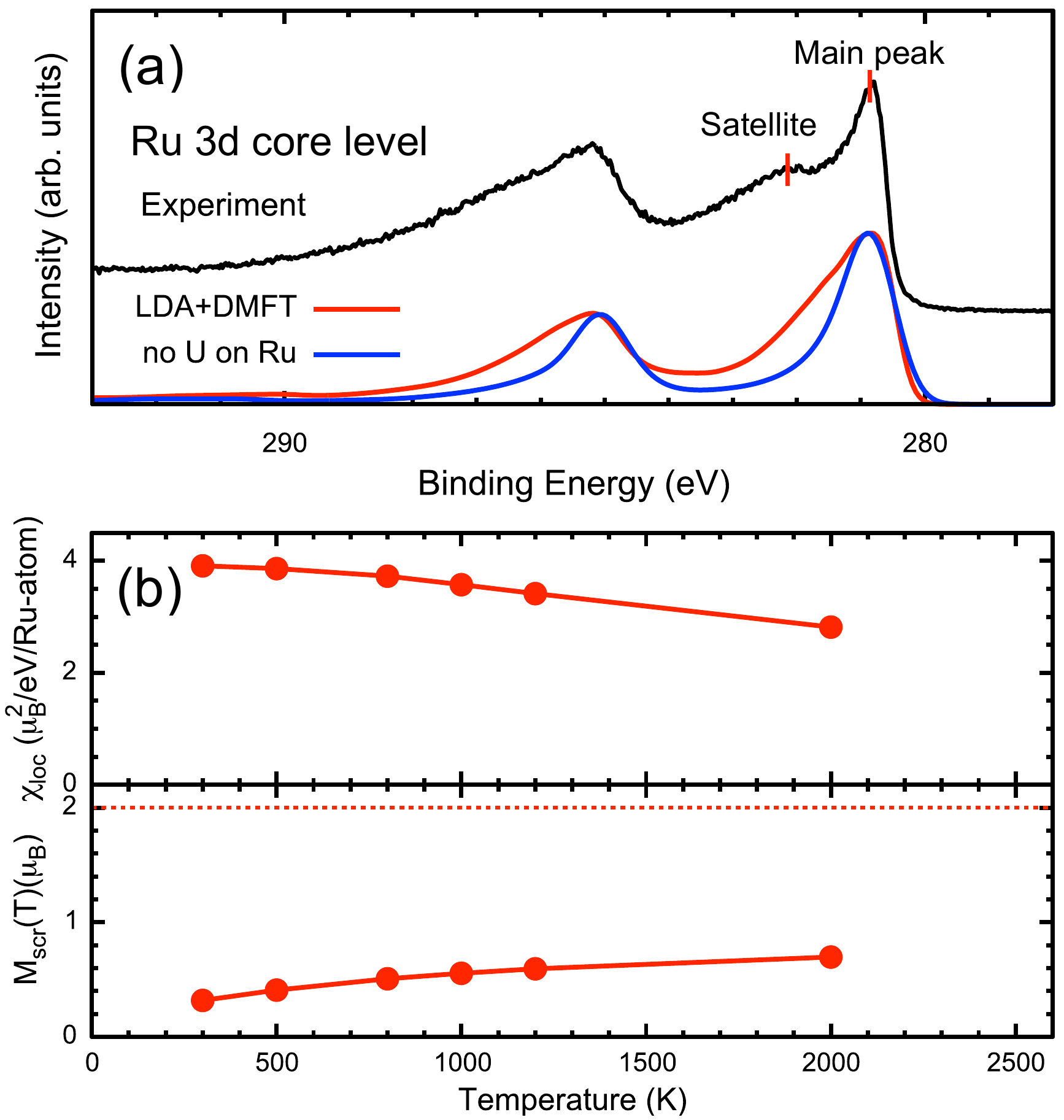}
	\caption{(a) Experimental Ru $3d$ core-level photoemission spectrum of CaCu$_3$Ru$_4$O$_{12}$
		(black line) and theoretical spectra from the LDA+DMFT calculation with $U_{\rm Ru}$ = 3.1~eV 
		(red line) and $U_{\rm Ru}$ = 0.0~eV (blue line).
		\textcolor{black}{(b) (top) Local spin susceptibility and (bottom) screened spin moment $M_{\rm scr}(T)=\sqrt{T\chi_{\rm loc}(T)}$~\cite{krapek12,hariki17ru} on the Ru site computed by the LDA+DMFT method. The horizontal dashed line indicates the atomic value of the Ru$^{4+}$ ion.}
		}
	\label{fig:Ru3d_core}
\end{figure}

\textcolor{black}{The parameters $\mu_{\rm dc}^{\rm Cu}$ and $\mu_{\rm dc}^{\rm Ru}$ of our theory (Appendix~C) are fixed 
to reproduce
(i) positions of the peaks A--E in the valence 
   valence photoemission spectra of Fig.~\ref{fig:vb_nw},
(ii) positions of the low-energy peak at 0.07--0.08~eV above, Fig.~\ref{fig:div}, and
   around 0.08~eV below~\cite{Liu20} the Fermi energy,
(iii) the splitting of the Cu 2$p_{3/2}$ XPS main line, see Fig.~\ref{fig:core},
  and (iv) observation of a shoulder in Ru 3$d$ core-level XPS spectrum, see Fig.~\ref{fig:Ru3d_core}.
}

%%%%%%%%%%%%%%%%%%%%%%%%%%%%%%%%%%%%%%%%%%%%%%%%%%%%%%%%%%%%%%%%%%%
\begin{figure}[t]
	\includegraphics[width=0.98\columnwidth]{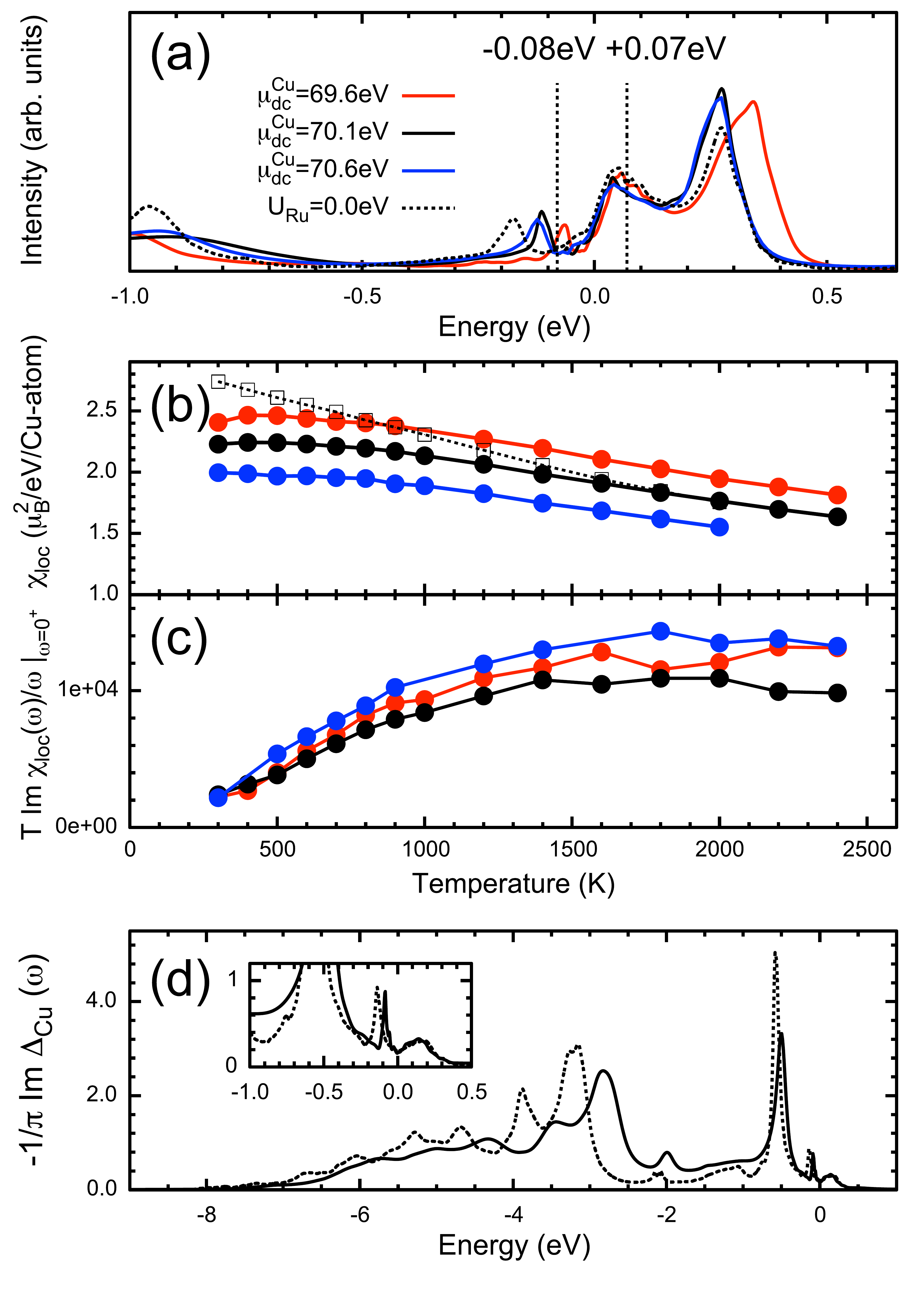}
	\caption{\textcolor{black}{(a) The Cu $xy$ spectral intensities in LDA+DMFT results for different double counting  $\mu^{\rm Cu}_{\rm dc}$ values. The positions of the low-energy peak in the experimental PES data are indicated by vertical dashed lines. The calculation is performed at 300~K. (b) Local susceptibility $\chi_{\rm loc}(T)$, (c) dynamical spin susceptibility $\chi_{\rm loc}(\omega={+0})$ at the Cu site. \textcolor{black}{Here the model with all Ru 4d states is employed. The dashed lines in panels (a) and (b) show the results obtained with $U_{\rm Ru}=0.0$~eV.
	(d) Hybridization densities $-1/\pi \textrm{Im}\Delta(\omega)$ of the Cu $xy$ orbital in LDA+DMFT with $U_{\rm Ru}=3.1$~eV (solid) and with $U_{\rm Ru}=0.0$~eV (dashed). The inset shows the hybridization densities near $E_F$.}}}
	\label{fig:lowe_app}
\end{figure}
%%%%%%%%%%%%%%%%%%%%%%%%%%%%%%%%%%%%%%%%%%%%%%%%%%%%%%%%%%%%%%%%%%%

%%%%%%%%%%%%%%%%%%%%%%%%%%%%%%%%%%%%%%%%%%%%%%%%%%%%%%%%%%%%%%%%%%%
%%%%%%%%%%%%%%%%%%%%%%%%%%%%%%%%%%%%%%%%%%%%%%%%%%%%%%%%%%%%%%%%%%%

\textcolor{black}{Before analysing $\mu_{\rm dc}^{\rm Cu}$ and $\mu_{\rm dc}^{\rm Ru}$ we briefly discuss why we need to include the \textcolor{black}{electron-electron} interaction on the Ru site \cite{Cao2008,Kikugawa2009,Liu2018}. To this end we have compared calculations with 
$U_{\rm Ru}=3.1$~eV %parameter on the Ru site 
from previous DFT studies for Ru oxides (with the same formal Ru valence as in CCRO)~\cite{gorelov10,pchelkina}
and with $U_{\rm Ru}=0.0$~eV. Fig.~\ref{fig:Ru3d_core}a shows that finite $U_{\rm Ru}$ is needed to yield the shoulder/satellite feature (iv) 
in the Ru 3$d$ core-level XPS spectra. However, finite $U_{\rm Ru}$ affects also the behavior of the Cu $xy$ state, \textcolor{black}{see dashed lines in Figs.~\ref{fig:lowe_app}ab}, through
dynamical renormalization (band narrowing) of Ru bands, which modifies the environment (hybridization density) of Cu, \textcolor{black}{see Fig.~\ref{fig:lowe_app}d}.
With $U_{\rm Ru}=0.0$~eV we could not find $\mu_{\rm dc}^{\rm Cu}$ to fulfill (i--iii).
}
\textcolor{black}{We would like to note that the Ru spin response is far from the Curie form despite the presence of correlations in the Ru $4d$ shell, see Fig.~\ref{fig:Ru3d_core}b. Correspondingly, also the screened (effective) Ru moment is substantially smaller than the atomic value ($\sim2\mu_B$) for a Ru$^{4+}$ ion.}

%%%%%%%%%%%%%%%%%%%%%%%%%%%%%%%%%%%%%%%%%%%%%%%%%%%%%%%%%%%%%%%%%%%
\begin{figure}[h]
	\includegraphics[width=0.95\columnwidth]{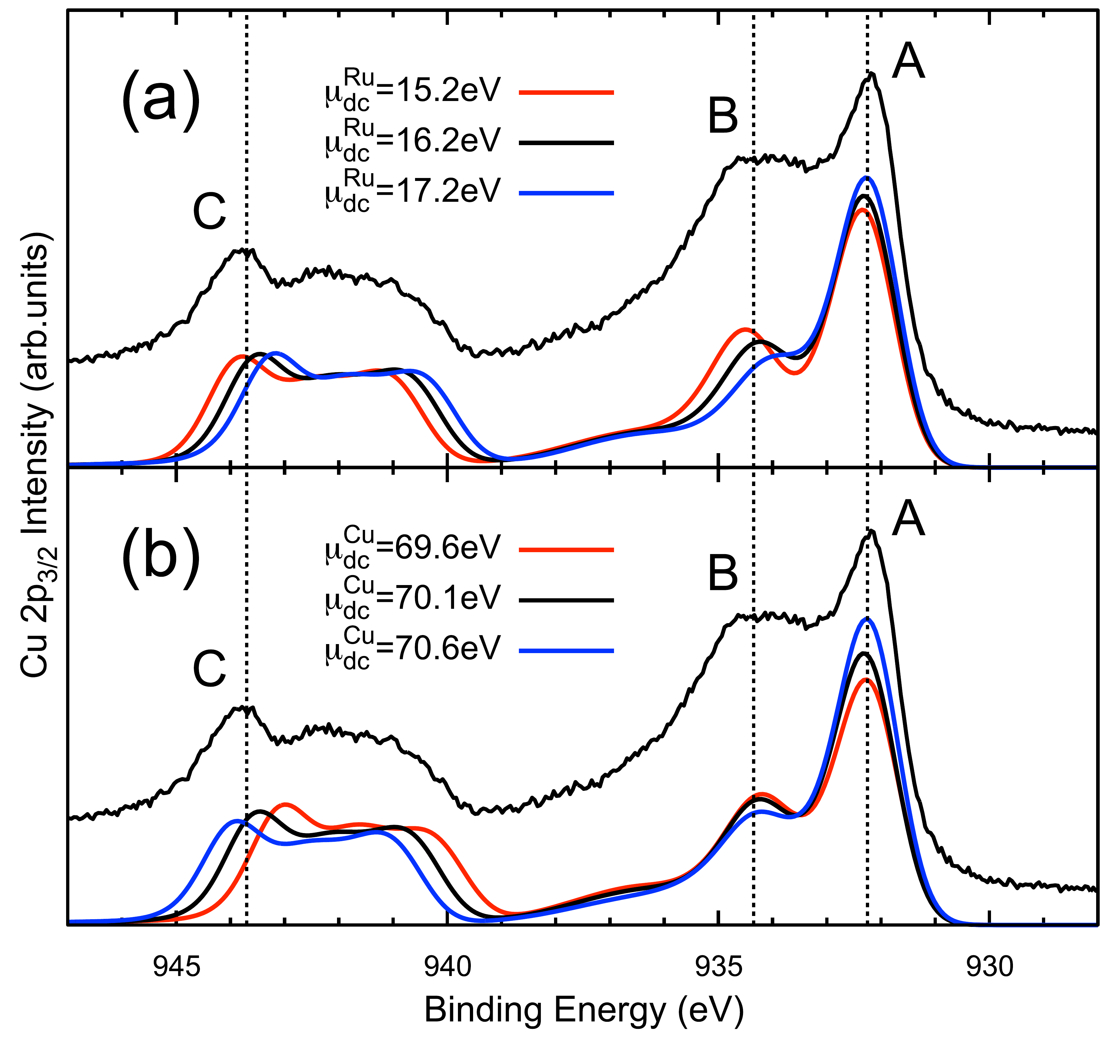}
	\caption{\textcolor{black}{(a) Cu 2$p_{3/2}$ core-level photoemission spectra calculated by the LDA+DMFT method with selected $\mu^{\rm Ru}_{\rm dc}$ values (with $\mu^{\rm Cu}_{\rm dc}=70.1$~eV). 
	(b) The spectra by LDA+DMFT with selected $\mu^{\rm Cu}_{\rm dc}$ values (with $\mu^{\rm Ru}_{\rm dc}=16.2$~eV).
	The experimental spectrum is shown for a comparison.}}
	\label{fig:cu_muru}
\end{figure}
%%%%%%%%%%%%%%%%%%%%%%%%%%%%%%%%%%%%%%%%%%%%%%%%%%%%%%%%%%%%%%%%%%%

%%%%%%%%%%%%%%%%%%%%%%%%%%%%%%%%%%%%%%%%%%%%%%%%%%%%%%%%%%%%%%%%%%%
\begin{figure}[h]
	\includegraphics[width=0.95\columnwidth]{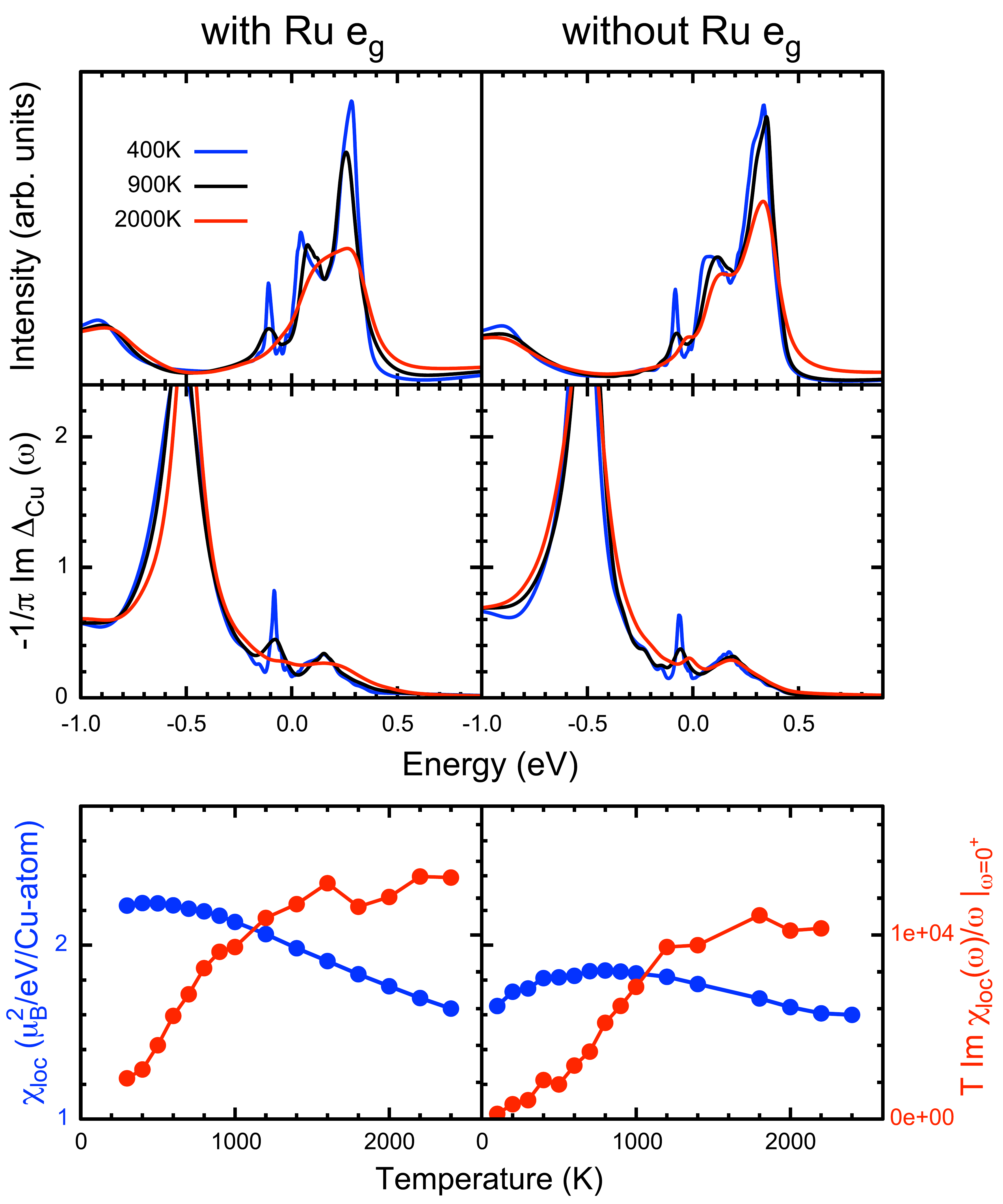}
	\caption{\textcolor{black}{LDA+DMFT results for the models (left) with and (right) without the Ru 4d $e_g$ states at selected temperatures. (top) Cu $xy$ spectral intensities, (middle) the hybridization densities of the Cu $xy$ orbitals, and (bottom) local susceptibility $\chi_{\rm loc}(T)$ and the dynamical spin susceptibility $\chi_{\rm loc}(\omega={+0})$ at the Cu site. $\mu^{\rm Cu}_{\rm dc}=70.1$~eV is employed in the two models, and $\mu_{\rm dc}^{\rm Ru}=16.2$~eV ($\mu_{\rm dc}^{\rm Ru}=13.4$~eV) for the model with (without) Ru 4$d$ $e_g$ states.}}
	\label{app_loc}
\end{figure}
%%%%%%%%%%%%%%%%%%%%%%%%%%%%%%%%%%%%%%%%%%%%%%%%%%%%%%%%%%%%%%%%%%%

\textcolor{black}{ 
The value of $\mu^{\rm Cu}_{\rm dc}$ affects the positions of the Cu $d^8$ satellite, feature $E$, and
feature B in Fig.\ref{fig:vb_nw} as well as the low-energy peaks (ii). All of these are well reproduced
by its chosen value of
$\mu^{\rm Cu}_{\rm dc}=70.1$~eV, see Fig.~\ref{fig:lowe_app}a \textcolor{black}{and Fig.~\ref{fig:vb_nw}}. 
\textcolor{black}{The small screened spin susceptibility due to the Kondo screening is found around the optimal $\mu^{\rm Cu}_{\rm dc}$ value, see Figs.~\ref{fig:lowe_app}bc.}
The $\mu^{\rm Cu}_{\rm dc}$ also affects the Cu--O charge-transfer energy and thus the splitting
between the main line and charge-transfer satellite in Cu 2$p$ XPS, see Fig.~\ref{fig:cu_muru}b.
On the other hand, the behavior of Cu, including variation of $\mu_{\rm dc}^{\rm Cu}$, has minor effect on its hybridization function, except for the peak just below Fermi energy. As a result the splitting
of Cu 2$p$ XPS main line (A,B) is independent of $\mu_{\rm dc}^{\rm Cu}$, see Fig.~\ref{fig:cu_muru}b.}

\textcolor{black}{The same is not true for $\mu^{\rm Ru}_{\rm dc}$. Affecting primarily the Ru--O charge-transfer energy, its variation modifies the Cu $3d$ $xy$ hybridization function around Fermi level, which has a sizable effect on the splitting in the Cu 2$p$ XPS main line (A,B), see Fig.~\ref{fig:cu_muru}a. We can thus use (iii) together with the Ru features in the global spectrum (i) to establish $\mu^{\rm Ru}_{\rm dc}=16.2$~eV.}

\textcolor{black}{Finally, we discuss the relationship of models with and without Ru $4d$ $e_g$ states. While $\mu^{\rm Cu}_{\rm dc}$ is the same for the two models, $\mu^{\rm Ru}_{\rm dc}$ must be different since the models include interaction between all Ru $4d$ orbitals in one case and between the $t_{2g}$ only in the other one. 
The matching $\mu^{\rm Ru}_{\rm dc}$ for the model without Ru $4d$ $e_g$ states is 13.2~eV. Both models yield the very similar temperature dependencies of the Cu spectral functions as well as hybridization densities and local susceptibilities, as shown in Fig.~\ref{app_loc}.}

\end{document}